# Multiphase Magnetic Systems: Measurement and Simulation

Yue Cao,[1] Mostafa Ahmadzadeh,[1] Ke Xu,[1] Brad Dodrill,[2] John S. McCloy[1,3,*]

1. Materials Science & Engineering Program, Washington State University, Pullman, WA, 99164, USA

2. Lakeshore Cryotronics, Westerville, OH, 43082, USA

3. School of Mechanical & Materials Engineering, Washington State University, Pullman, WA, 99164, USA

*john.mccloy@wsu.edu; +1-509-335-7796

## ABSTRACT

Multiphase magnetic systems are common in nature and are increasingly being recognized in technical applications. One characterization method which has shown great promise for determining separate and collective effects of multiphase magnetic systems is first order reversal curves (FORCs). Several examples are given of FORC patterns which provide distinguishing evidence of multiple phases. In parallel, a visualization method for understanding multiphase magnetic interaction is given, which allocates Preisach magnetic elements as an input 'Preisach hysteron distribution pattern' (PHDP) to enable simulation of different 'wasp-waisted' magnetic behaviors. These simulated systems allow reproduction of different major hysteresis loop, FORC pattern, and switching field distributions of real systems and parameterized theoretical systems. The experimental FORC measurements and FORC diagrams of four commercially obtained magnetic materials, particularly those sold as nanopowders, shows that these materials are often not phase pure. They exhibit complex hysteresis behaviors that are not predictable based on relative phase fraction obtained by characterization methods such as diffraction. These multiphase materials, consisting of various fractions of $BaFe_{12}O_{19}$, $\varepsilon\text{-}Fe_2O_3$, and $\gamma\text{-}Fe_2O_3$, are discussed.

**Keywords: Wasp-waisted hysteresis; FORC; barium hexaferrite; $\varepsilon\text{-}Fe_2O_3$**

I. INTRODUCTION

Even though the ideal characteristics of hysteresis in perfect single magnetic domains is generally accepted, it is still a challenge to interpret the hysteresis of real materials, as it depends on particle size, shape, and distribution, as well as stress, defects, and impurities. The hysteresis of magnetic mixtures is even more complex, as interactions between non-dilute phases distort the simple loop shape. However, the parameters extracted from major hysteresis loop – for instance, coercivity – are not very efficient at showing the interactions of magnetic phases in a multiphase system.

First order reversal curve (FORC) analysis is one technique which has developed rapidly in the last two decades,[1] providing a powerful tool for observing magnetic switching contributed by different magnetic phases. FORC is now applied to understand various hysteretic behaviors in metal-insulator transitions,[2] ferroelectricity,[3] magnetocalorics,[4] and perpendicular magnetic recording media.[5] The FORC technique can also be employed to distinguish the different magnetic signals in multiphase systems[4,6] to investigate subtle magnetic features caused by interaction of multiple magnetic phases.

In this paper, a series of multiphase magnetic nanopowders are investigated experimentally using major hysteresis loops and FORC diagrams and these systems are also modeled using 'Preisach hysteron distribution patterns' (PHDPs) based on the classical Preisach model (CPM) to aid in descriptive understanding of the reduction of coercivity, or 'wasp-waistedness' features, observed in major loops of multiphase materials. The FORC diagrams of them generally indicate the existence of a low coercivity phase, a high coercivity phase, and a coupling region, which can all be simulated using PHDPs. Together, the simulations and measurements provide a coherent picture of magnetic interactions in a 'wasp-waisted' system,[7,8] and they confirm the utility of the CPM for describing such systems.

## II. SIMULATIONS AND EXPERIMENTAL METHODS

### A. Theory and Simulation

As the basis of First Order Reversal Curves (FORC), the Preisach model[9] is a typical model of hysteresis, and it is a simple and straightforward description of magnetic switching. It has been developed with increasing complexity to refine its description of magnetic behaviors observed under different conditions,[10,11] but the classic Preisach model (CPM) is used here as a useful phenomenological approach to understanding highly interacting multiphase magnetic materials (see Supplementary Material for more details on simulations and relationship to the Preisach model). In the proposed embodiment of the CPM, all hysterons are distributed onto a two-dimensional (2D) coordinate system to build a 'Preisach hysteron distribution pattern' (PHDP) (Fig. S-1(b)). The magnetization of the PHDP at a given external field is calculated via summing the magnetization of all hysterons, as the overall magnetization of a real material could be decomposed into a series of these hysterons. Since each hysteron has magnetization of nominally +1 or -1, the number of hysterons in a simulation, while phenomenological on a first order, represents both the mole fraction of a phase and the relative magnetization of that phase. In other words, if a phase has higher molar magnetization, a mole of said phase would be represented by more hysterons than a different phase with a smaller magnetization.

The FORC technique captures hysteretic features on multiple reversal curves, and the magnetization of every data point on each reversal curve is determined by the reversal field ($H_r$) and the magnetic field ($H$). Thus, for each data point on a given FORC, its reversal field $H_r$ is equivalent to the Preisach switching field $a$, and its magnetic field $H$ represents the Preisach switching field $b$. Then a FORC diagram is plotted with contours of FORC distribution density ($H_r$, $H$) with coercivity $H_c(H_r, H)$ as the x-axis and interaction $H_u(H_r, H)$ as the y-axis. In this paper, the FORC distribution density $\rho$ of PHDP is obtained via least squares fitting a second-order polynomial function using Matlab 2014b[12,13] and the corresponding FORC diagrams are processed in Origin 9.1. More detail regarding the PHDP and FORC technique have been discussed more thoroughly in previous publications[10,14] and in the Supplementary Material.

## B. Sample selection and FORC measurements

To investigate the two-component PHDP of a real sample to compare to simulation, FORC data was acquired on a series of commercial materials, all purchased as M-type barium hexaferrite ($BaFe_{12}O_{19}$). $BaFe_{12}O_{19}$ (Ba-M) is a commercial and competitive permanent magnet[15,16] as magnetic recording material,[17,18] and millimeter wave absorber.[19] The micrometer-sized sized $BaFe_{12}O_{19}$ powders were obtained from advanced ferrite technology (AFT) GmbH and nanometer-sized materials were obtained from Aldrich.[20] For the purposes of this paper, these samples have been denoted "Micro" (large particle $BaFe_{12}O_{19}$), and "Nano-1", "Nano-2", and "Nano-3" for the three lots of nanopowders purchased as "$BaFe_{12}O_{19}$." The particle size,[21] crystal phase identification by X-ray diffraction (XRD),[20] temperature-dependent magnetization,[21] and major hysteresis loops[20] have been previously assessed. These materials were assessed to have multiple crystalline phases, most of which were not $BaFe_{12}O_{19}$, and major loops exhibited characteristic shapes indicative of interacting phases with different magnetic properties. The XRD and major loops up to 50 kOe are reviewed and replotted in the Fig S-4 and Fig S-3.

Most samples considered had very high coercivity and high saturation field. Multiple attempts were made to capture the whole hysteresis behavior in FORC (see Supplementary Material). Ultimately, all samples were measured on a Lakeshore 7400 vibrating sample magnetometer (VSM) with saturation magnetic field $H_{sat}$ of ±33 kOe, which is considerably higher than standard VSMs used for FORC measurements. The specific parameters, including the field steps, the number of FORCs and the smoothing factors, for each sample are listed in Table I. Note that in multiple cases, the loops in FORC appeared to saturate in lower field instruments, but previous major loop behavior observed up to 50 kOe could not be reproduced (see Supplementary Material). This indicates an important caveat in FORC measurements of high concentrations of very high coercivity materials which are typically not encountered in geologic contexts. All FORC diagrams were processed using the VARIFORC function in FORCinel,[22,23] which is written in Igor (V2.02 in IGOR Pro7, WaveMetrics, Portland, OR). In processing the FORC data, the lower edge artifact was removed[24] prior to plotting the FORC diagram.

**Table I**. Final FORC parameters. $H_{u1}$ and $H_{u2}$ are the limits for the y-axis (interaction or bias) on the FORC diagram. $H_{c1}$ and $H_{c2}$ are the limits for the x-axis (coercivity axis) on the FORC diagram. $H_{Cal}$ and HSat are the fields used for calibration or saturation, respectively, applied at the end of each FORC. $H_{Ncr}$ is field step between reversal fields and $N_{Forc}$ is the number of FORCs.

| Sample | $H_{u1}$ (kOe) | $H_{u2}$ (kOe) | $H_{c1}$ (kOe) | $H_{c2}$ (kOe) | $H_{Cal}$ (kOe) | $H_{Sat}$ (kOe) | $H_{Ncr}$ (Oe) | $N_{Forc}$ |
|---|---|---|---|---|---|---|---|---|
| Micro | -1.5 | 1.5 | 0 | 10 | 11.845 | 20 | 89.65 | 150 |
| Nano-1, 2, and 3 | -1.5 | 1.5 | 0 | 28 | 29.5 | 33 | 208.84 | 150 |

## III. RESULTS

### A. FORC diagrams and interpretation

#### 1. Micro sample

The experimentally obtained raw FORCs and calculated FORC diagram of the 'Micro' sample are shown in Fig. 1. It is apparent that a small but more compact peak appears at low coercivity (~100 Oe) and a widely distributed peak appears at high coercivity (~3 kOe). However, a significant 'wasp-waistedness' was not clear in the previously measured major hysteresis loop, probably due to a large field step in the original major loop measurement, which was 0.5 kOe (see Fig. S-3). Note that the FORC measurements used significantly smaller step sizes which did resolve multiple features. The concentric distribution at high coercivity suggests single domain (SD) particles,[25,26] and the large vertical spread of the high coercivity component suggests the interaction between these SD particles.[1,27] The bias to negative interaction ($H_u$) suggests mean field magnetizing interactions,[28] which would be expected from packed hexagonal plate[21] crystals. The distribution of low coercivity component also showed spread along the $H_u$ axis, but the spread gradually decreases with increasing coercivity, which suggests pseudo-single domain (PSD) particles.[29] The designation of PSD is a transition region between SD and multi domain (MD), and the coercivity of PSD and MD particles decrease with increasing particle size.[30,31]

Though XRD suggested only $BaFe_{12}O_{19}$ phase,[21] it is possible that peaks of magnetite ($Fe_3O_4$) and/or maghemite (γ-$Fe_2O_3$) could be hidden in the pattern (see Supplementary Material). Maghemite is the fully oxidized equivalent of magnetite with the same crystal structure. Its typical reported saturation magnetization is ~74 emu/g, whereas that of magnetite is ~92 emu/g.[31] The SD critical grain size of maghemite is ~60 nm and it is ~50-84 nm for magnetite,[32] although these values can depend on particle shape.[33] Almeida et al[34] have observed almost identical FORC behavior for magnetite and maghemite with similar particle size. Therefore, their similar behaviors makes it difficult to exclusively assign the low-$H_C$ component to one of these phases. It is also possible, though unlikely, that the low coercivity peak is large domain $BaFe_{12}O_{19}$ of the PSD size. However, the coercivity of $BaFe_{12}O_{19}$ does not appear to drop for particles at least 1 μm in size,[35] which is about the observed particle size for this material,[21] though some larger particles are possible. The high coercivity component is most likely single domain $BaFe_{12}O_{19}$ particles, since the peak in coercivity is ~3 kOe, which is well in line with previous reports on the coercivity of this material from ~50 nm to >1 μm (2.5 – 6 kOe).[20,21,35]

2. *Nano-1 sample*

The FORC diagram of 'Nano-1' (Fig. 2), on the other hand, shows two separated peaks at $H_c$ = ~0.2 kOe and $H_c$ = ~6.5 kOe and one obscure peak at $H_c$ = ~2 kOe. Again, the low coercivity component behaves as PSD, and the middle and high coercivity components are most likely two populations of SD particles. The coercivity of the middle peak is quite close to that of SD $BaFe_{12}O_{19}$ particles in the 'Micro' sample, and therefore, the middle coercivity component is probably SD $BaFe_{12}O_{19}$. The high coercivity component was initially assigned to SD ε-$Fe_2O_3$ nanoparticles; however, the reasonable coercivity of SD ε-$Fe_2O_3$ nanoparticles within similar particle size has been reported as ~15-20 kOe, which is much higher than the value observed.[36] The high coercivity component was finally determined as SD $BaFe_{12}O_{19}$ (but with a different particle size distribution than the middle peak) after comparing to the theoretical and experimental results.[35,36] We attempted to capture the signal of single domain $\varepsilon$-$Fe_2O_3$ nanoparticles by changing the intensity of contour, as XRD shows considerable amount of this phase, but still did not observe any peak beyond 6.5 kOe. The disappearance of the low-magnetization, high-coercivity component has been reported in some magnetic mixtures, especially in multi-component systems which show significant magnetization difference between two different magnetic phases. For example, 95% hematite ($M_s$ ~0.4 emu/g) was dominated by 5% magnetite ($M_s$ ~92 emu/g),[37] and the resulting FORC diagram only suggested the existence of magnetite phase at room temperature.[38,39] By comparison, $BaFe_{12}O_{19}$ has $M_s$ ~50-70 emu/g[21] and $\varepsilon$-$Fe_2O_3$ has $M_s$ ~15 emu/g.[36] Note that, the fact that there is no offset for the high coercivity component indicates those single domain nanoparticles are randomly oriented.[12] Corroborating evidence from the XRD result (see Supplementary Material), the low coercivity component is likely PSD $\gamma$-$Fe_2O_3$ or magnetite. The average particle size of 'Nano-1' is in the PSD transition region for $\gamma$-$Fe_2O_3$ and magnetite.[31]

3. *Nano-2 sample*

The major loop of 'Nano-2' shows the most significant 'wasp-waistedness' since the two observed distribution peaks ($H_c$ ~ 0.2 kOe and 24 kOe) are widely separated in its FORC diagram (Fig. 3). The XRD identified that the major phase in 'Nano-2' was $\varepsilon$-$Fe_2O_3$ phase, but could not rule out the possibility of small amounts of other Fe oxides, such as $Fe_3O_4$ or $\gamma$-$Fe_2O_3$. The FORC diagram shows a clear peak located at ~24 kOe, which is close to the peak value of coercivity of SD $\varepsilon$-$Fe_2O_3$ particles.[36] Note that, the primary particles size is 30 to 50 nm,[21] which is still in the SD range of $\varepsilon$-$Fe_2O_3$. The low coercivity component, on the other hand, could be larger PSD $\varepsilon$-$Fe_2O_3$ particles or PSD $Fe_3O_4$/ $\gamma$-$Fe_2O_3$. Since a small amount of $Fe_3O_4$/ $\gamma$-$Fe_2O_3$ can dominate over the FORC signal from $\varepsilon$-$Fe_2O_3$, due to their magnetization difference, the low coercivity component in the FORC diagram of 'Nano-2,' like the other samples, is likely $Fe_3O_4$/ $\gamma$-$Fe_2O_3$ particles.

*4. Nano-3 sample*

'Nano-3' shows a pinched major loop with low major loop coercivity, presumably due to high concentration of PSD γ-$Fe_2O_3$ particles (low coercivity and thin loop) in addition to SD ε-$Fe_2O_3$ particles. It is worth pointing out that the XRD patterns of γ-$Fe_2O_3$ and $Fe_3O_4$ are very similar, so again magnetite versus maghemite cannot be reliably distinguished with the current data. The FORC diagram of 'Nano-3' (Fig. 4) does not show multiple peaks, unless measured up to very high magnetic field. A subtle high coercivity peak then emerges at very high coercivity (~25 kOe), which suggests that this peak still belongs to single domain ε-$Fe_2O_3$ particles, with lower fraction than Nano-2, while the low coercivity component is assigned to PSD γ-$Fe_2O_3$/$Fe_3O_4$ particles which contribute more magnetization, and hence higher intensity. XRD showed that Nano-3 contains ~50% γ-$Fe_2O_3$/$Fe_3O_4$ phase which could contribute more magnetization, and eventually hide the signal of SD ε-$Fe_2O_3$ particles in its FORC diagram.

B. Simulations to match experimental FORC diagrams

In order to tailor further the resulting FORC diagrams for comparison with experimental diagrams, a phenomenological Preisach model was created using three different concentric ellipses/circles of hysterons to represent the possible phases described in Sec. III A. The concentric ellipses are used to describe $BaFe_{12}O_{19}$ phase(s) and ε-$Fe_2O_3$ phase(s), and the concentric circles are created for γ-$Fe_2O_3$ phase. In addition, to emphasize the difference between the magnetization of each phase, the total number of hysterons in a concentric ellipse/circle for different phases are intentionally designed. For example, one concentric ellipse for ε-$Fe_2O_3$ contains only 20 hysterons, but one concentric ellipse for $BaFe_{12}O_{19}$ consists of 120 hysterons due to its high magnetization. $BaFe_{12}O_{19}$ phase and ε-$Fe_2O_3$ phase are distinguishable by the long axis/short axis ratio. Since $BaFe_{12}O_{19}$ phases show wider spread along the $H_u$ axis in the experimental results, the long axis/short axis ratio of the $BaFe_{12}O_{19}$ phase is designed larger than that of the ε-$Fe_2O_3$ phase. All parameters of different phases are listed in Table II.

Table II. Summary of simulation parameters of $BaFe_{12}O_{19}$, $\varepsilon$-$Fe_2O_3$ and $\gamma$-$Fe_2O_3$ phase.

| Phase Name | # of hysterons in one concentric ellipse/circle | Long axis/short axis ratio | Color |
|---|---|---|---|
| $BaFe_{12}O_{19}$ | 100 | 4/3 | Green |
| $\varepsilon$-$Fe_2O_3$ | 20 | 2/1 | Blue |
| $\gamma$-$Fe_2O_3$ | 120 | N/A | Red |

Four different elliptical PHDPs were created to simulate the elongated distributions along the $H_c$ axis in the FORC diagram of Micro and Nano samples. All possible phases discussed in Sec. III A are represented by grouped concentric ellipses or circles of hysterons. The centers of the elliptical distribution are allocated on the diagonal, and their major axes are parallel to the diagonal to obtain symmetric distributions. Previous study showed that the center of hysteron distribution determined the coercivity; therefore, the same phase of particles that show size-dependent coercivity are exhibited by locating the center of ellipses/circles at various position on the PHDP diagonal. It is also valid that a phase present in a larger amount is modeled by multiple concentric distributions in this simulation. To slow down the magnetization switching at the remnant state, all concentric ellipses and circles are distributed on a uniform matrix where the hysterons are placed with distance of 0.1 horizontally and vertically.

The simulated FORC dataset of each PHDPs consist of 40 FORCs, and the magnetization of each data point on every reversal curve is calculated at a defined step of 0.05 until the magnetic field H returns to saturation again. It was found that simulation of 40 FORCs was sufficient to present all the major features of typically obtained experimental FORC diagrams, and additional FORCs only slightly improved resolution. It should be recognized that either increasing the number of FORCs or decreasing the step of FORC will change the quality of FORC diagrams,[26] resulting in longer collecting time in a FORC measurement and longer processing time to extract the FORC diagram. Additionally, the quality of a FORC diagram is also affected by the smoothing factor (SF).[12] In this modeling, SF=2 is chosen to map FORC diagrams in better resolution and higher quality. Smaller

SF will induce more noise and larger SF will cut off partial concentric circles at high coercivity region.

The simulated phases and corresponding simulation parameters in four elliptical PHDPs are summarized in Table III. Note that the low coercivity phase was deliberately changed for M1 ($BaFe_{12}O_{19}$), M2 ($\gamma$-$Fe_2O_3$), and M3 ($\varepsilon$-$Fe_2O_3$) to illustrate the fact that the FORC diagram is not strongly affected by the choice of this phase. It has been confirmed in selected cases that replacement of $\gamma$-$Fe_2O_3$ for $\varepsilon$-$Fe_2O_3$ in M3, for instance, gives a FORC pattern and FORC diagram if the total number of hysterons for the group of concentric circles or ellipses is kept constant.

**Table III**. Summary of elliptical PHDP for Ba-M; for all simulations, the matrix step size is 0.1, the number of FORCs is 40, the step is 0.05, and the SF=2.

| PHDP # | Phase 1 | Centroid 1 | Phase 2 | Centroid 2 | Phase 3 | Centroid 3 | Phase 4 | Centroid 4 | # of concentric distribution of each phase | Similar sample |
|---|---|---|---|---|---|---|---|---|---|---|
| M1 | $BaFe_{12}O_{19}$ | [-0.05, 0.05] | $BaFe_{12}O_{19}$ | [-0.3, 0.3] | N/A | N/A | N/A | N/A | 1;5 | Micro |
| M2 | $\gamma$-$Fe_2O_3$ | [-0.05, 0.05] | $BaFe_{12}O_{19}$ | [-0.2, 0.2] | $BaFe_{12}O_{19}$ | [-0.4, 0.4] | $\varepsilon$-$Fe_2O_3$ | [-0.2, 0.2] | 1;1;5;1 | Nano-1 |
| M3 | $\varepsilon$-$Fe_2O_3$ | [-0.05, 0.05] | $\varepsilon$-$Fe_2O_3$ | [-0.6, 0.6] | N/A | N/A | N/A | N/A | 4;4 | Nnano-2 |
| M4 | $\gamma$-$Fe_2O_3$ | [-0.05, 0.05] | $\varepsilon$-$Fe_2O_3$ | [-0.6, 0.6] | N/A | N/A | N/A | N/A | 1;1 | Nano-3 |

III. DISCUSSION

Barium hexaferrite written as $BaFe_{12}O_{19}$ can also be written as $BaO \cdot 6Fe_2O_3$. Thus, one can imagine that incomplete reaction with BaO during synthesis could result in $Fe_2O_3$ phases, and the BaO could be present as an amorphous phase and be undetectable with XRD. It has been suggested that under certain synthesis conditions, the smallest particles of $Fe_2O_3$ tend toward the $\gamma$-$Fe_2O_3$ phase, while the $\varepsilon$-$Fe_2O_3$ phase is stable from ~3-8 to ~30 nm.[40,41] Small $\gamma$-$Fe_2O_3$ particles <4 nm are superparamagnetic, while those 4-8 nm have coercivitites less $\leq$ 2.1 kOe.[41] While not probably in the size-range of PSD maghemite, these low coercivities could account for the low coercivity components observed in all the samples. It has also been reported that $\varepsilon$-$Fe_2O_3$ can form in the presence of $Ba^{2+}$ if the Fe/Ba ratio is 10-20.[42] Hematite, $\alpha$-$Fe_2O_3$, can also form in this series as particles of ε-Fe$_2$O$_3$ become larger than ~30 nm.[40,41] Apparently, an uncontrolled commercial synthesis process resulted in the nanopowders with one or more Fe$_2$O$_3$ phases rather than BaFe$_{12}$O$_{19}$.

FORC is readily able to distinguish these different phases, provided that the maximum applied field in the FORC sequence is high enough. In a number of cases, a lower field resulted in an incomplete diagram being collected, and the highest coercivity would have easily been missed if the major loops had not already been measured on a high field instrument.[20] These FORCs which do not achieve high enough field are not representative of the whole magnetic behavior of the system, but rather only the low coercivity components, much like minor loops.

Use of phenomenological simulation allowed the recreation of the expected FORC and FORC diagram behavior in the presence of multiple phases with different magnetizations and coercivities. There are more robust simulations for FORC diagrams based on micromagnetic simulations, for example,[24] but are restricted to SD particles. FORC measurements provide a quick way for identifying characteristic phases in a mixed sample. When coercivities are much different than one another, especially when the high coercivity phase has low magnetization, the remanence and hence FORC diagram can be dominated by the high magnetization phase, such as in the case of magnetite/hematite mixtures.[38] While it is tempting to consider a simple superposition of the signatures separate phases, it is clear that this does not always happen and considerable interaction between the particles takes place, particularly in highly interacting, non-dilute conditions such as considered here.

## IV. SUMMARY AND CONCLUSION

Four samples of commercially obtained "barium hexaferrite" were identified by First Order Reversal Curves at room temperature and up to high fields. In two cases, high coercivity ε-Fe$_2$O$_3$ could be identified by FORC when it was suggested by XRD, but not in the third case where roughly half the sample was γ-Fe$_2$O$_3$ (or Fe$_3$O$_4$) a high magnetization and low coercivity phase. Even in the apparently crystallographically pure micron-sized BaFe$_{12}$O$_{19}$, strong evidence of low coercivity phases, possibly pseudo-single domain structures, were found at coercivities <2 kOe. High coercivity phases $BaFe_{12}O_{19}$ (2-5 kOe) and $\varepsilon$-$Fe_2O_3$ (20-26 kOe), had broad coercivity distributions shown by elongated positive regions on the FORC diagram along the $H_c$ axis. Most FORC diagrams show evidence of strong interparticle magnetic bias by broadening in the $H_u$ interaction axis. In the case of the 'micro' sample, the centroid $BaFe_{12}O_{19}$ was clearly shifted to negative $H_u$, suggesting magnetizing mean field interactions. Simple phenomenological Preisach modeling allowed reproduction of the main features of the FORC diagrams, and showed that simple mixing of the phases does not necessarily produce a peak for each phase.

## SUPPLEMENTARY MATERIAL

See the Supplementary Material for details on simulation theory and methods, X-ray diffraction data and discussion, complete FORC and FORC diagrams of all iterations of the experiments, and isothermal remnant magnetization (IRM) data for Nano-1.

## ACKNOWLEDGEMENT

This research was funded by the U.S. Department of Energy in support of the Nuclear Energy Enabling Technologies– Reactor Materials (NEET-3) program. The authors thank the anonymous reviewer and Neil Dilley for help improving the manuscript.

FIGURE CAPTIONS

**Fig. 1.** (a) Experimental FORCs (every 2$^{nd}$ FORC displayed) with major loop and (b) corresponding FORC diagram of 'Micro' at room temperature with VARIFORC smoothing factor $S_{c0} = S_{b1} = 5$, $S_{c1} = S_{b1} = 7$. The unit of the FORC density intensity scale bar is emu/kg/Oe$^2$. Inset shows close-up near $H_c$~0.

**Fig. 2.** (a) Experimental FORCs (every 2$^{nd}$ FORC displayed) with major loop and (b) corresponding FORC diagram of 'Nano-1' at room temperature with VARIFORC smoothing factor $S_{c0} = S_{b1} = 5$, $S_{c1} = S_{b1} = 7$. The unit of the FORC density intensity scale bar is emu/kg/Oe$^2$. Inset shows close-up near $H_c$~0.

**Fig. 3.** (a) Experimental FORCs (every 5$^{th}$ FORC displayed) with major loop and (b) corresponding FORC diagram of 'Nano-3' at room temperature with VARIFORC smoothing factor $S_{c0} = S_{b1} = 5$, $S_{c1} = S_{b1} = 7$. The unit of the FORC density intensity scale bar is emu/kg/Oe$^2$. Inset shows close-up near $H_c$~0.

**Fig. 4.** (a) Experimental FORCs (every FORC displayed) with major loop and (b) corresponding FORC diagram of 'Nano-3' at room temperature with VARIFORC smoothing factor $S_{c0} = S_{b1} = 5$, $S_{c1} = S_{b1} = 7$. The unit of the FORC density intensity scale bar is emu/kg/Oe$^2$. Inset shows close-up near $H_c$~0 and near max field. In the main FORC diagram, contour of the high coercivity component is drawn in, while in the inset, the color-sale was adjusted to make this component visible; this inset color scale does not correspond to main figure.

**Fig. 5.** (a-d) Simulated elliptical PHDPs and (e-h) corresponding FORC and (i-l) FORC diagrams for EP1, EP2, EP3 and EP4 with SF=2. $BaFe_{12}O_{19}$, $\gamma\text{-}Fe_2O_3$ and $\varepsilon\text{-}Fe_2O_3$ phases are represented by green, red and blue, respectively. Black spots are matrix hysterons.

FIGURES

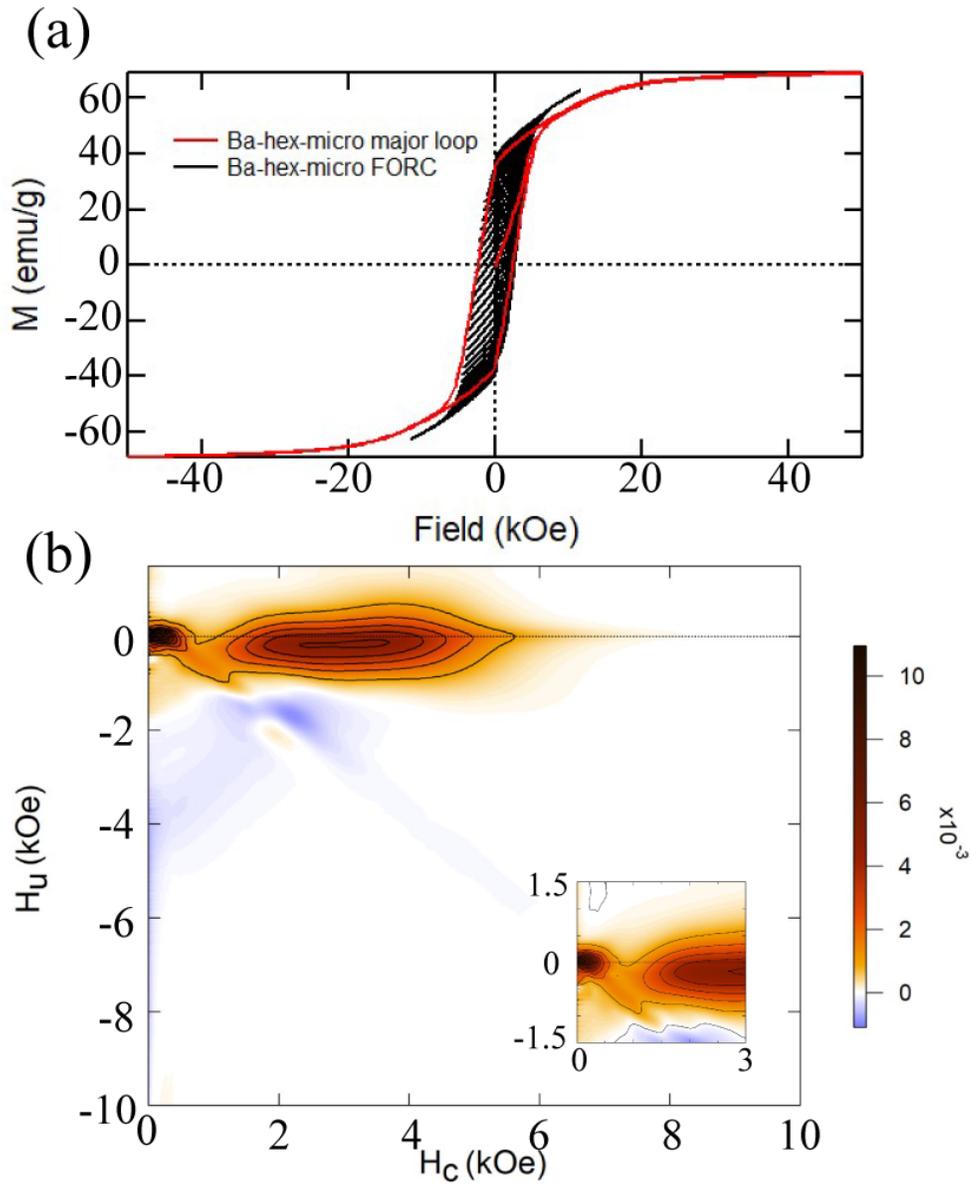

**Fig. 1.** (a) Experimental FORCs (every 2$^{nd}$ FORC displayed) with major loop and (b) corresponding FORC diagram of 'Micro' at room temperature with VARIFORC smoothing factor $S_{c0} = S_{b1} = 5$, $S_{c1} = S_{b1} = 7$. The unit of the FORC density intensity scale bar is emu/kg/Oe$^2$. Inset shows close-up near $H_c \sim 0$.

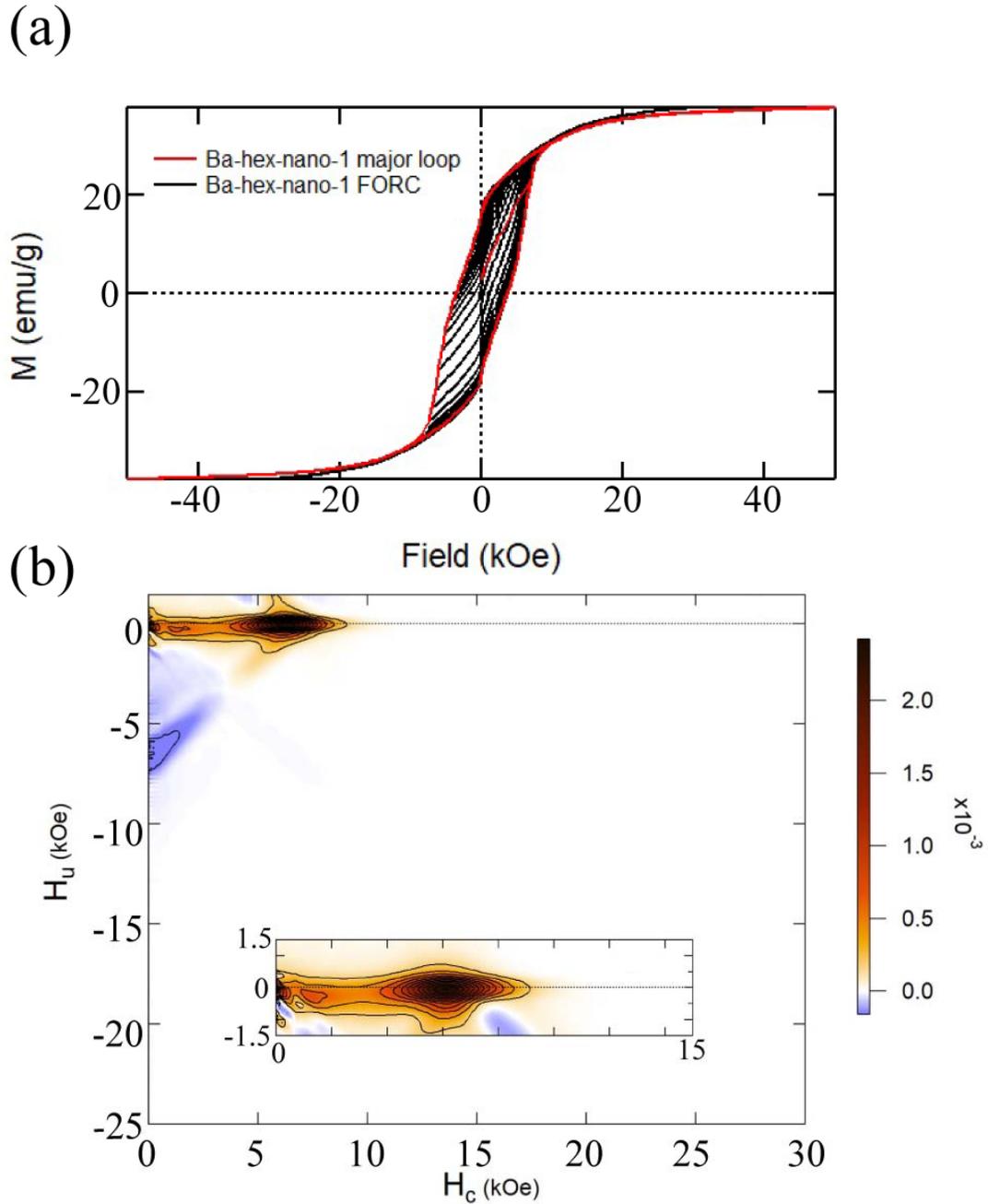

**Fig. 2.** (a) Experimental FORCs (every 2$^{nd}$ FORC displayed) with major loop and (b) corresponding FORC diagram of 'Nano-1' at room temperature with VARIFORC smoothing factor $S_{c0} = S_{b1} = 5$, $S_{c1} = S_{b1} = 7$. The unit of the FORC density intensity scale bar is emu/kg/Oe$^2$. Inset shows close-up near $H_c \sim 0$.

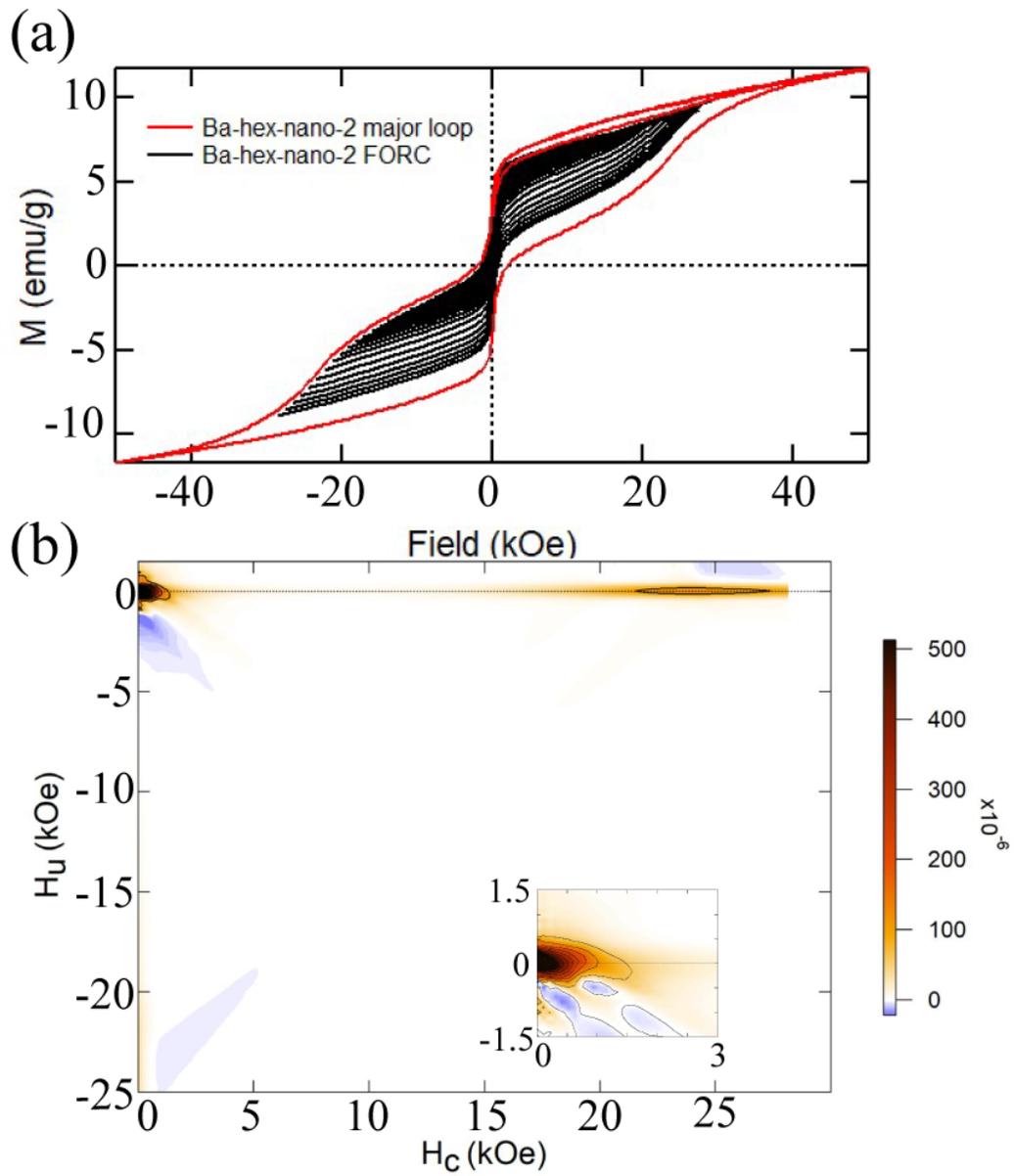

**Fig. 3.** (a) Experimental FORCs (every 5$^{th}$ FORC displayed) with major loop and (b) corresponding FORC diagram of 'Nano-3' at room temperature with VARIFORC smoothing factor $S_{c0} = S_{b1} = 5$, $S_{c1} = S_{b1} = 7$. The unit of the FORC density intensity scale bar is emu/kg/Oe$^2$. Inset shows close-up near $H_c \sim 0$.

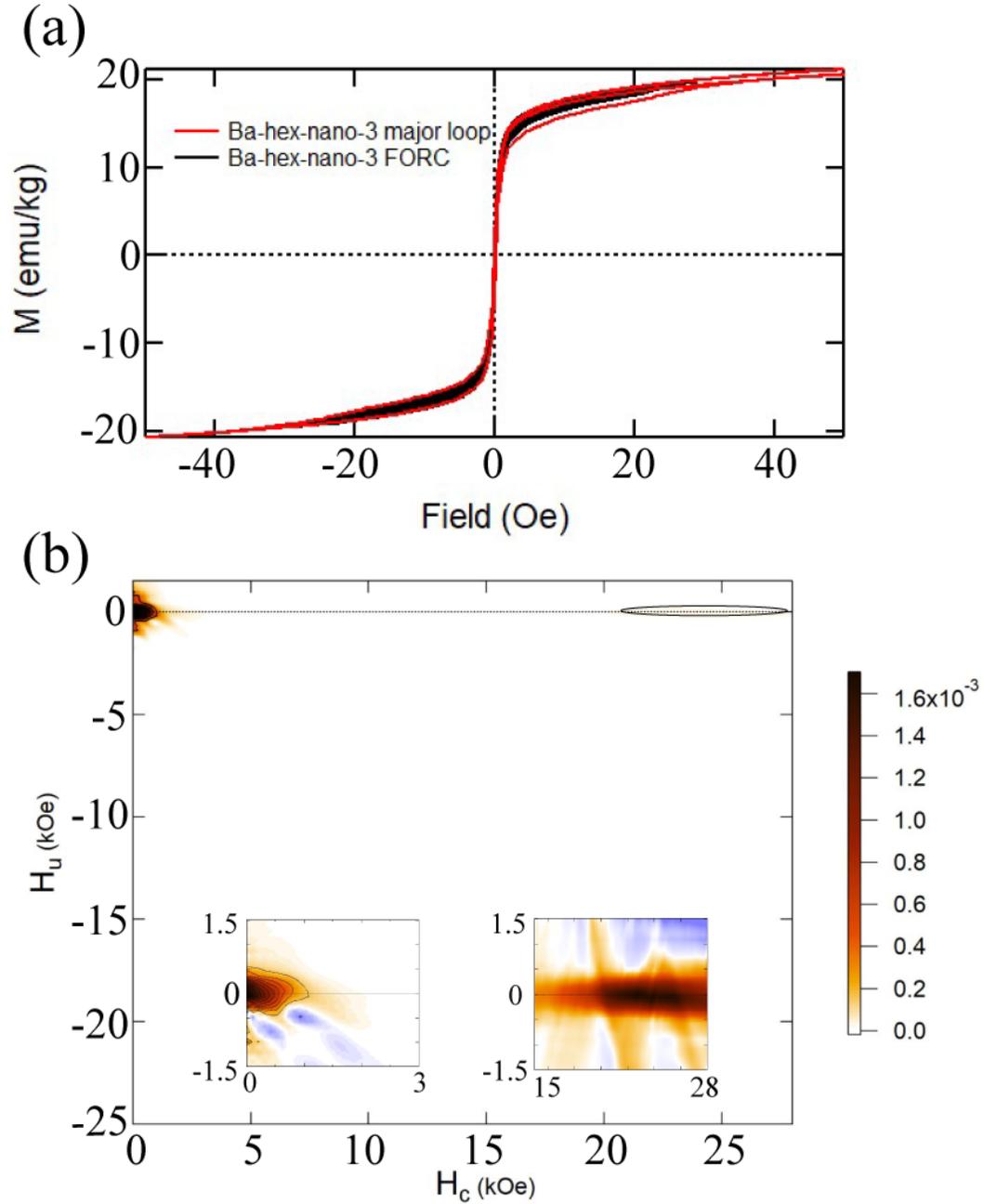

**Fig. 4.** (a) Experimental FORCs (every FORC displayed) with major loop and (b) corresponding FORC diagram of 'Nano-3' at room temperature with VARIFORC smoothing factor $S_{c0} = S_{b1} = 5$, $S_{c1} = S_{b1} = 7$. The unit of the FORC density intensity scale bar is emu/kg/Oe$^2$. Inset shows close-up near $H_c \sim 0$ and near max field. In the main FORC diagram, contour of the high coercivity component is drawn in, while in the inset, the color-sale was adjusted to make this component visible; this inset color scale does not correspond to main figure.

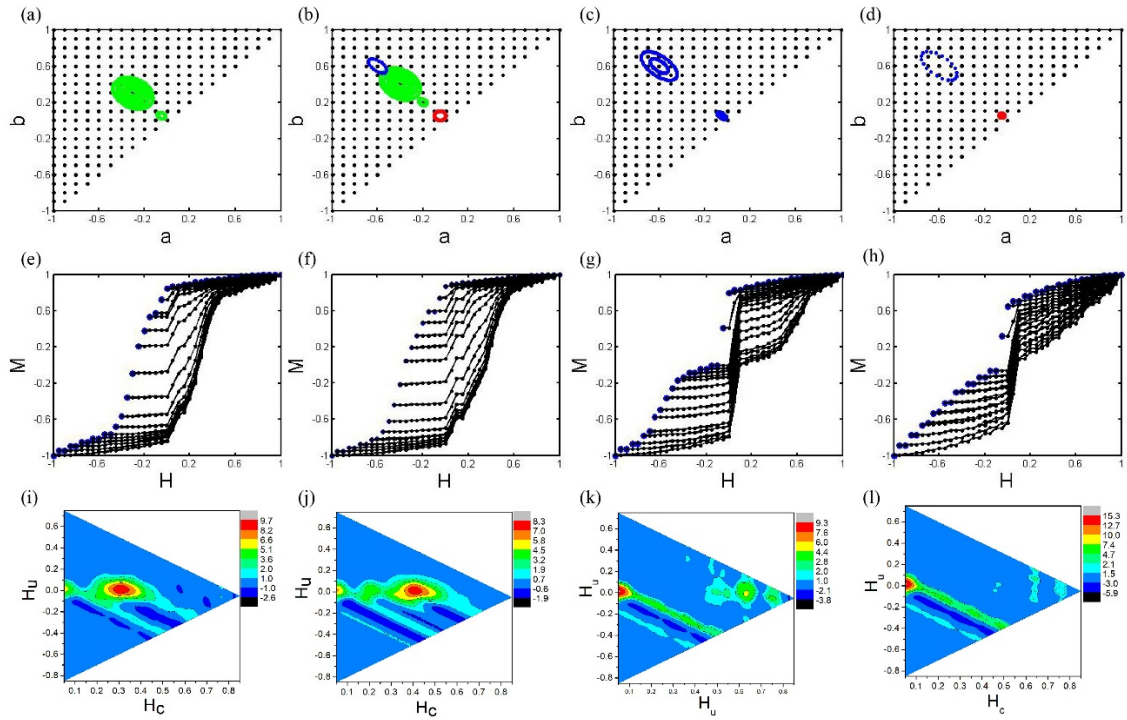

**Fig. 5.** (a-d) Simulated elliptical PHDPs and (e-h) corresponding FORC and (i-l) FORC diagrams for M1, M2, M3 and M4 with SF=2. $BaFe_{12}O_{19}$, $\gamma$-$Fe_2O_3$ and $\varepsilon$-$Fe_2O_3$ phases are represented by green, red and blue, respectively. Black spots are matrix hysterons.

# SUPPLEMENTARY MATERIAL
# Multiphase Magnetic Systems: Measurement and Simulation

Yue Cao,[1] Mostafa Ahmadzadeh,[1] Ke Xu,[1] Brad Dodrill,[2] John S. McCloy[1,3]

*1. Materials Science & Engineering Program, Washington State University, Pullman, WA, 99164, USA*
*2. Lakeshore Cryotronics, Westerville, OH, 43082, USA*
*3. School of Mechanical & Materials Engineering, Washington State University, Pullman, WA, 99164, USA*

- 23 pages
- 4 tables
- 16 figures

I. DETAILS ON SIMULATIONS

A. Theory

As the basis of First Order Reversal Curves (FORC), the Preisach model[1] is a typical model of hysteresis, and it is a simple and straightforward description of magnetic switching. It has been developed with increasing complexity to refine its description of magnetic behaviors observed under different conditions.[2,3] However, despite the growing diversity of the Preisach model manifestations, none satisfactorily applies to all situations, so the utility of the classic Preisach model (CPM) may still be considerable given its simplicity.

In the CPM, one 'Preisach hysteron' is a mathematical magnetic element which has two independent switching fields ($a$ and $b$) and two states of magnetization (1 and -1). The shape of hysteresis of one hysteron is perfectly rectangular in the CPM.[2] The magnetization ($M$) of each hysteron is determined by switching fields and external magnetic field ($H$):

$$M(H) = \begin{cases} +1 & (H \geq b) \\ -1 & (H \leq a) \\ x & (b > H > a) \end{cases} \quad (1)$$

where $x$ is -1 when the hysteresis loop is ascending and +1 when it is descending. The configuration for a hysteresis loop of a single hysteron is shown in Fig. S-1(a).

In the proposed embodiment of the CPM, all hysterons are distributed onto a two-dimensional (2D) coordinate system to build a 'Preisach hysteron distribution pattern' (PHDP) (Fig. S-1(b)). It should be noted that this is a very general embodiment of a distribution, whereas other more specific ones, such as lognormal, Lorentzian, or Gaussian, have been previously suggested.[3-5] The magnetization of the PHDP at a given external field is calculated via summing the magnetization of all hysterons, as the overall magnetization of a real material could be decomposed into a series of these hysterons. Since each hysteron has magnetization of nominally +1 or -1, the number of hysterons in a simulation, while phenomenological on a first order, represents both the mole fraction of a phase and the relative magnetization of that phase. In other words, if a phase has higher molar magnetization, a mole of said phase would represent more hysterons than a different phase with a smaller magnetization.

For every hysteron in the PHDP, its coercivity ($H_c$) and interaction ($H_u$) (see Fig. S-1(a)) are defined as:[2]

$$H_c(a,b) = \frac{b-a}{2} \qquad H_u(a,b) = \frac{b+a}{2} \quad (2)$$

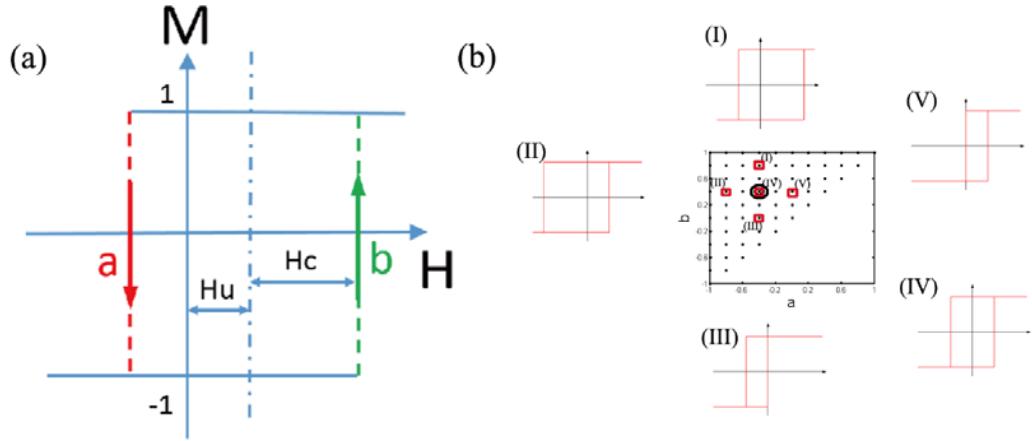

**Fig. S-1**. (a) Schematic of a single magnetic hysteron element. (b) Simulated PHDP (middle) and 5 hysterons with switching fields (*a,b*) as (I) (-0.4, 0.8); (II) (-0.8, 0.4); (III) (-0.4, 0); (IV) (-0.4, 0.4) and (V) (0, 0.4).

B. Classic Preisach model of multi-phase materials

Fig. S-2(a-d) shows selected simulated PHDPs using the CPM, and the simulation parameters of more PHDPs are summarized in Table S-I. Each PHDP contains two parts: 1) a uniform matrix and 2) series of concentric ellipses (with a circle being the symmetric case) with various centroids and radii. The radius of the outside concentric circle is taken as 0.05, 0.1 or 0.25 to obtain various hysteretic switching behavior [2]. More concentric circles are placed by gradually decreasing the radius with a gap of 0.05 until the centroid is reached. Each concentric circle consists of 80 evenly distributed hysterons, and it is distributed on the uniform matrix where hysterons are placed with distance of 0.2 horizontally and vertically. In this paper, the effect from the position and concentration of the concentric circles is primarily discussed, since it impacts the hysteresis of PHDPs more significantly than the matrix.[2] Simulations P1 and P2 have only one distribution centroid, but P3, P4, P5, and P6 are designed to contain two centroids, representing either a single magnetic phase with a bimodal size distribution or two different magnetic phases. The impact of a shift from a circular to elliptical distribution is discussed in Sec. III B in the main paper.

The simulated FORC dataset of each PHDPs consist of 40 FORCs, and the magnetization of each data point on every reversal curve is calculated at a defined step of 0.05 until the magnetic field H returns to saturation again. It was found that simulation of 40 FORCs was sufficient to present all the major features of typically obtained experimental FORC diagrams, and additional FORCs only slightly improved resolution. It should be recognized that either increasing the number of FORCs or decreasing the step of FORC will change the quality of FORC diagrams,[6] resulting in longer collecting time in a FORC measurement and longer processing time to extract the FORC diagram. Additionally, the quality of a FORC diagram is also affected by the smoothing factor (SF).[7] In this modeling, SF=2 is chosen to map FORC diagrams in better resolution and higher quality. Smaller SF will induce more noise and larger SF will cut off partial concentric circles at high coercivity region.

The corresponding FORCs of each PHDP are simulated and shown in Fig. S-2(e-h). The coercivity $|H_c|$ of the major hysteresis loop is identical for P1 and P2 since they have the same centroid of concentric circles.[2] The coercivity of the PHDP is located at an intermediate point between the centroids for two-component PHDPs. Another feature which can be readily appreciated is that the radius of concentric circles controls the susceptibility $\chi_c$ of the hysteresis loop at the coercivity. It has been shown that increasing the radius of concentric circles (P1 versus P2) decreases the sharpness of the hysteresis as well as the susceptibility $\chi_c$.[2] However, it can be observed that adding another series of concentric circles at another centroid can also decrease the susceptibility (e.g., P1 versus P3). Additionally, the radius (P5 versus P6, in Table S-I) and the centroid (P3 versus P5) of the added concentric circles can lead to either an increase or a decrease of the susceptibility.

The third observation is that moving the two centroids away from each other can result in an observable "wasp-waistedness" in the hysteresis (e.g., P3 and P4). The wasp-waisted hysteresis is generally caused by: 1) two phases or materials which have different magnetic behaviors[8,9] or 2) an inhomogeneously distributed single magnetic phase.[10] It is worth mentioning that the wasp-waisted hysteresis is not only attributed to multiple magnetic phase coupling. For example, the shape asymmetry[11] or vortex state magnetization reversal[12] can result in wasp-waistedness as well.

Note that a pair of tails are observed in simulated FORC diagrams, which are numerical artifacts from the data points that cannot be perfectly fitted the polynomial FORC function. These features are also referred to as the 'lower edge artifact' that is removable using commercial FORC processing software (e.g. FORCinel).[13]

The FORC diagrams of P3 and P4, representing two-components PHDPs, (Fig. S-2(c-d)) contain two peaks which correspond to the two centroids of PHDPs. One interesting feature is that two peaks are extracted in the FORC diagrams of P3, while its major hysteresis loop does not show obvious wasp-waistedness. This is because these two centroids in P3 are very close to each other, demonstrating that a FORC diagram can be more sensitive than other standard magnetic measurements.

**Table S-I.** Summary of PHDP of CPM; for all simulations, the matrix step size is 0.2, the number of FORCs is 40, the step is 0.05, and the SF=2. Here radius is the radius of the outermost circle, $\chi_c$ is the susceptibility (i.e., slope dM/dH) at the coercivity, $|H_c|$ is the absolute value of the coercivity. See text for description of the distribution density.

| PHDP # | Centroid 1 | Radius 1 | Distribution density 1 | Centroid 2 | Radius 2 | Distribution density 2 | $\chi_c$ | $|H_c|$ |
|---|---|---|---|---|---|---|---|---|
| P1 | [-0.4, 0.4] | 0.1 | 2.42 | N/A | N/A | N/A | 7.39 | 0.40 |
| P2 | [-0.4, 0.4] | 0.25 | 6.06 | N/A | N/A | N/A | 5.43 | 0.40 |
| P3 | [-0.2, 0.2] | 0.1 | 2.42 | [-0.4, 0.4] | 0.1 | 2.42 | 2.85 | 0.30 |
| P4 | [-0.1, 0.1] | 0.1 | 2.42 | [-0.5, 0.5] | 0.25 | 6.06 | 4.06 | 0.45 |
| P5 | [-0.2, 0.2] | 0.1 | 2.42 | [-0.5, 0.5] | 0.1 | 2.42 | 1.76 | 0.35 |
| P6 | [-0.2, 0.2] | 0.1 | 2.42 | [-0.5, 0.5] | 0.25 | 6.06 | 4.05 | 0.45 |

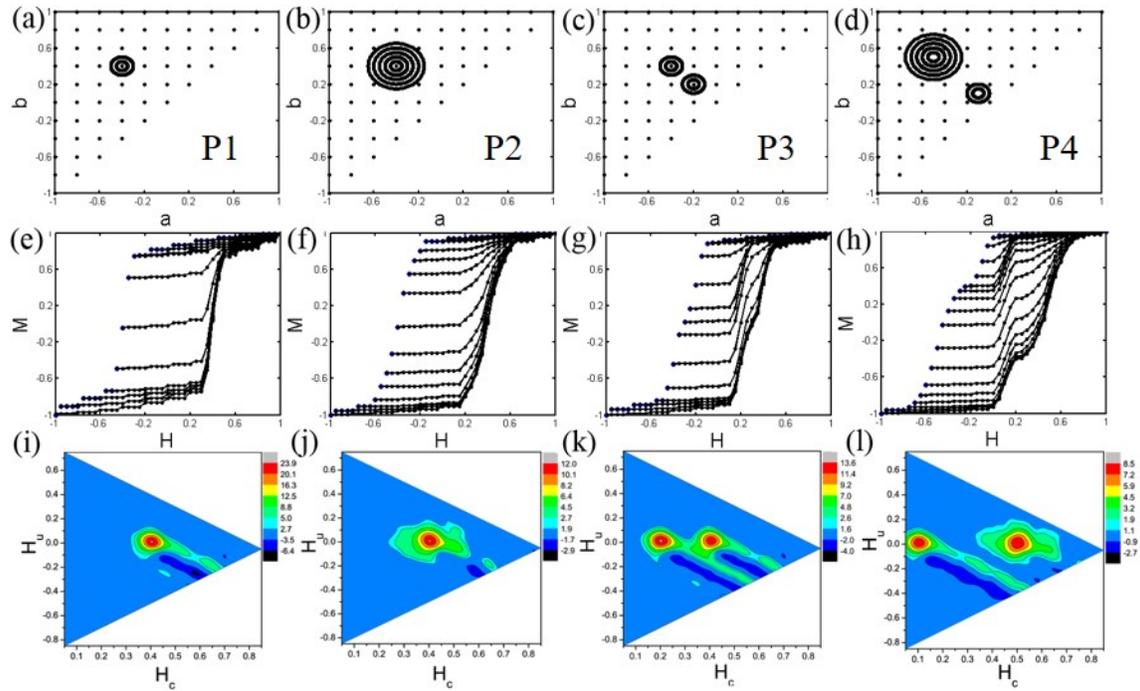

**Fig. S-2.** (a-d) Simulated PHDPs and (e-h) corresponding FORC and (i-l) FORC diagrams for P1, P2, P3 and P4 with SF=2.

The simulated elliptical PHDPs, the resulting FORCs, and the corresponding FORC diagrams are shown in Fig. S-3 and parameters in Table S-II. The FORC diagrams of EP1, EP2 and EP3 show two peaks spread at the low coercivity region and high coercivity region, representing the two centroids in the elliptical PHDPs. Apparently, the distance between the centroids decides the 'wasp-waistedness' of the hysteresis loop. For example, the hysteresis loop of EP1 closely resembles the loop of 'Micro,' which exhibit insignificant 'wasp-waistedness.'

**Table S-II**. Summary of elliptical PHDP; for all simulations, the matrix step size is 0.1, the number of FORCs is 40, the step is 0.05, and the SF=2.

| PHDP # | Centroid 1 (ellipse) | Long axis | Short axis | Distribution density 1 | Centroid 2 (circle) | Radius 2 | Distribution density 2 | $\chi_c$ | $|H_c|$ |
|---|---|---|---|---|---|---|---|---|---|
| EP1 | [-0.3, 0.3] | 0.3 | 0.1 | 3.03 | [-0.05, 0.05] | 0.05 | 0.61 | 4.61 | 0.3 |
| EP2 | [-0.4, 0.4] | 0.3 | 0.1 | 3.03 | [-0.05, 0.05] | 0.05 | 0.61 | 4.86 | 0.37 |
| EP3 | [-0.5, 0.5] | 0.3 | 0.1 | 3.03 | [-0.05, 0.05] | 0.05 | 0.61 | 5.12 | 0.45 |
| EP4 | [-0.05, 0.05] | 0.1 | 0.03 | 3.03 | [-0.05, 0.05] | 0.05 | 0.61 | 12.74 | 0.13 |

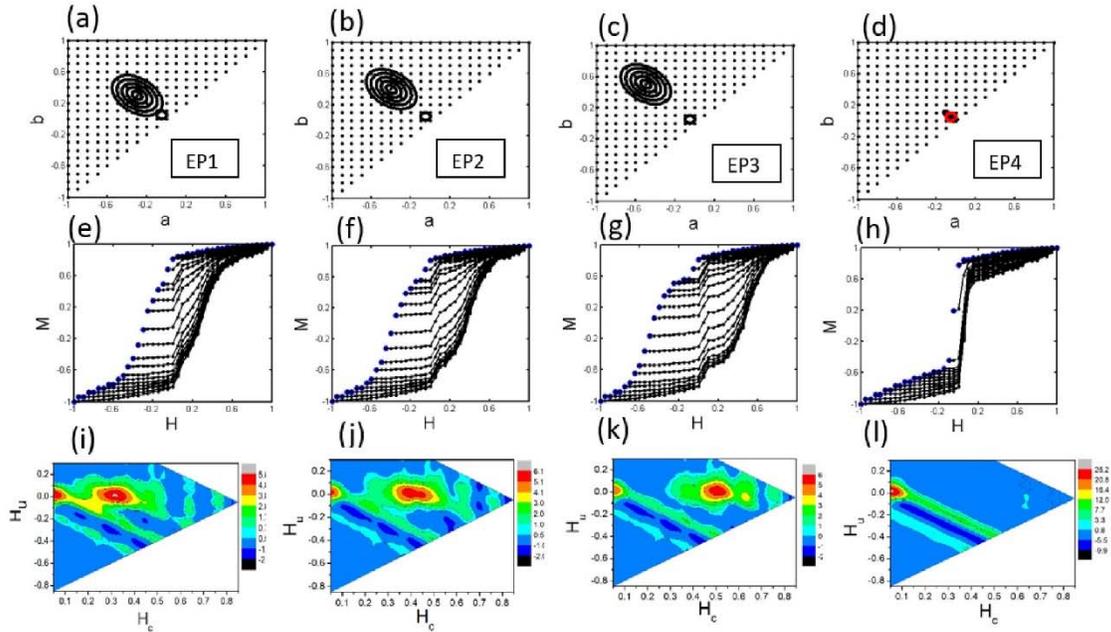

**Fig. S-3.** (a-d) Simulated elliptical PHDPs and (e-h) corresponding FORC and (i-l) FORC diagrams for EP1, EP2, EP3 and EP4 with SF=2. Red circle in (d) represents the circular distribution, and the elliptical distribution (black) overlays the red circle.

## II. DETAILS ON EXPERIMENTS

### A. Initial characterization of commercial powders

#### i. Magnetic data

Previously collected major loops are shown in Fig. S-4. Note the strong wasp-waistedness in some samples, particularly "nano-2" and "nano-3" samples.

Major loop saturation magnetization $M_s$ and coercivity $H_c$ are consistent with those measured previously, indicating that the phases had not magnetically changed in the period between these two measurements (~6 years). Experimental FORCs were collected on several different instruments. Magnetic measurements for 'barium hexaferrite' systems were collected at room temperature.

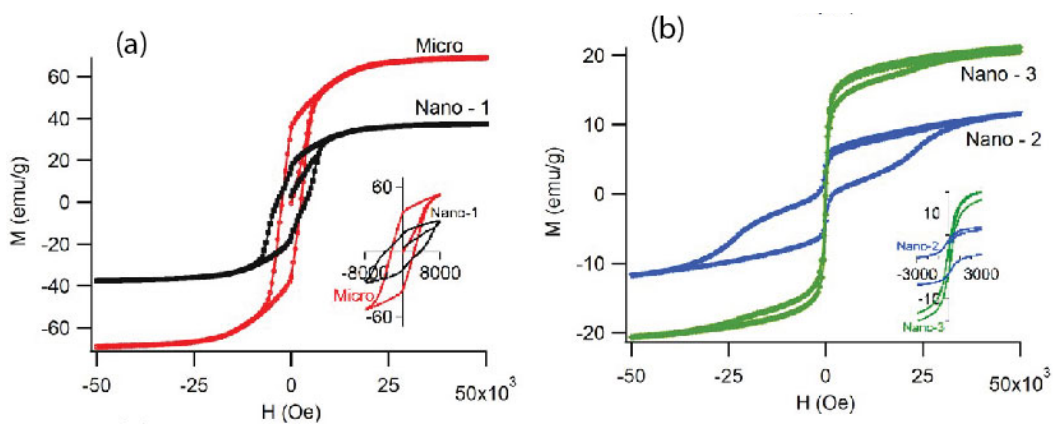

**Fig. S-4.** Major hysteresis loops from powders of (a) and (b) 'barium hexaferrite' systems at room temperature

#### ii. X-ray diffraction data

A summary of the crystallographic phase determination from X-ray diffraction (XRD) obtained via Rietveld refinement is shown in Table S-III, based on previously analyzed data.[14] Note that some samples of "barium hexaferrite" did not contain any measurable quantity of this phase.

**Table S-III**. Details on studied nanoparticle systems

| Name | Description | XRD Phase ID (vol%)[14] |
|---|---|---|
| Ba-hex-micro | ~ 1 μm particles of barium hexaferrite[30,31] | 100% $BaFe_{12}O_{19}$ |
| Ba-hex-nano-1 | ~60 nm particles of mixed Fe-oxides[14,15] | 56% $BaFe_{12}O_{19}$, 44% $\varepsilon\text{-}Fe_2O_3$ |
| Ba-hex-nano-2 | ~60 nm particles of mixed Fe-oxides[14] | 100% $\varepsilon\text{-}Fe_2O_3$ |
| Ba-hex-nano-3 | ~60 nm particles of mixed Fe-oxides[14] | 51% $\gamma\text{-}Fe_2O_3$, 49% $\varepsilon\text{-}Fe_2O_3$ |

Given the first order reversal curve data (FORC) data obtained on these samples, a reanalysis of the XRD data was conducted.

Raw XRD spectra of the samples, along with the reference patterns of the identified phases are shown in Figure S-4. Ba-hex (ICSD-980066757) is the only phase that is detected in the Micro sample which shows intense sharp peaks. However, small amount of magnetite (ICSD-980158745) could be present since the main magnetite diffraction peak overlaps with Ba-hex peaks. Moreover, one should note that the diffraction patterns of magnetite and maghemite ($\gamma$-$Fe_2O_3$) are very similar, which makes it difficult to discriminate between the two phases in XRD spectra. Therefore, the pseudo-single domain (PSD) behavior in the corresponding FORC diagram is likely due to the presence of magnetite and/or maghemite with relatively large spontaneous magnetization.

The nano-crystalline nature of the Nano samples leads to the peak broadening observed in their diffraction patterns. XRD spectrum of Nano-1 powder reveals the presence of considerable amounts of other phases in addition to Ba-hex. The XRD analysis shows the characteristic peaks of $\varepsilon$-$Fe_2O_3$ phase (ICSD-980415250), which intrinsically has an extremely large coercive field. Minor fractions of magnetite/maghemite and hematite (ICSD-980161292) are likely to be present in this sample. The PSD signature in the FORC diagram can be assigned to magnetite/maghemite, while the single domain (SD) distributions from FORC are likely due to Ba-hex phase with distinct size distributions. Hematite's signature (if any) cannot be observed in the FORC, because FORC measurements are not able to reveal signals from hematite (which has very small spontaneous magnetization) when other high-$M_S$ phases are simultaneously present.[9] Similarly, due to relatively low magnetization of the $\varepsilon$-$Fe_2O_3$ phase, no sign of this phase is observed in the FORC diagram of Nano-1 sample, even at very high fields.

High-$H_C$ $\varepsilon$-$Fe_2O_3$ is the major phase identified in the XRD pattern of Nano-2 sample. This phase is responsible for the distribution peak with very high coercivity (~ 25 kOe) in the FORC diagram. The PSD component in the FORC diagram is likely due to the presence of magnetite/maghemite whose main diffraction peaks can be seen as low-intensity broad peaks in the XRD spectrum. It is worth mentioning that, despite the large fraction of $\varepsilon$-$Fe_2O_3$ phase in this sample, its distribution peak shows lower intensities in the FORC diagram than that of magnetite/maghemite due to its lower magnetization.

The major phase in the Nano-3 sample is magnetite/maghemite. However, XRD shows the presence of considerable $\varepsilon$-$Fe_2O_3$ as well. Therefore, the corresponding FORC diagram shows an obvious PSD behavior assigned to magnetite/maghemite, along with a subtle significantly high-$H_C$ SD component from $\varepsilon$-$Fe_2O_3$.

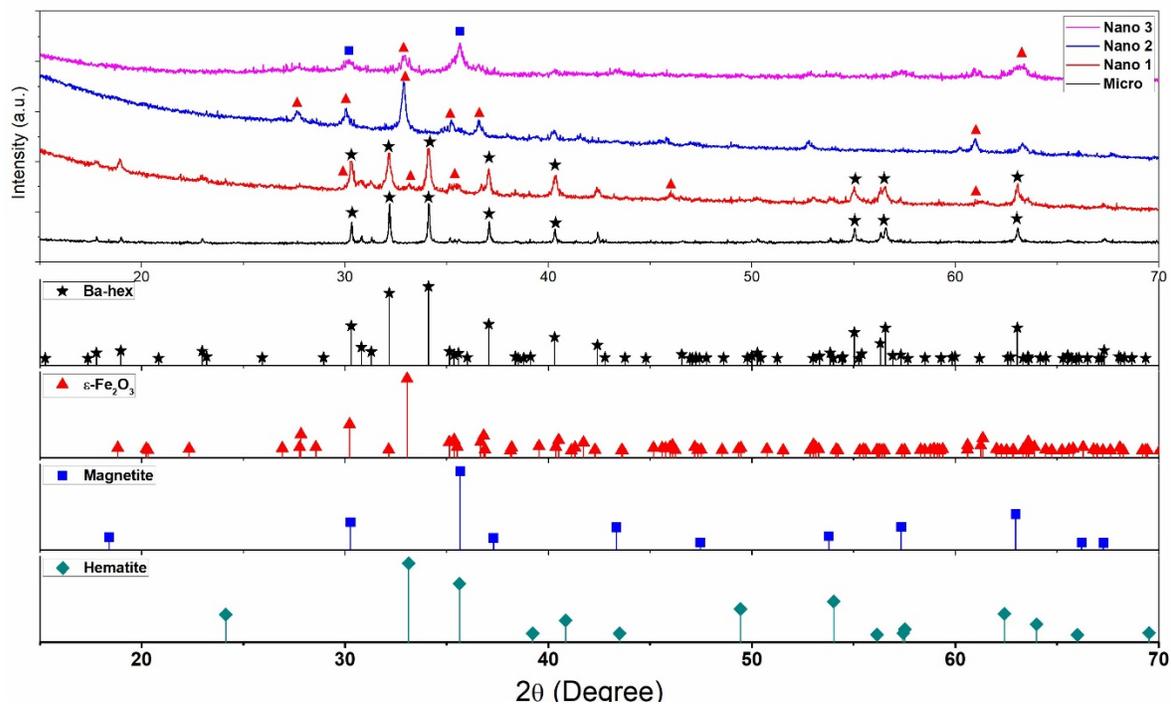

**Fig. S-5.** X-ray diffraction results for the four samples measured, along with labeled characteristic peaks and patterns for the potential phases. ICSD numbers for the phases are provided in the text.

B. Isothermal Remnant Magnetization (IRM) measurements

There are several different techniques to statistically "unmix" the measured magnetization curves, a procedure which is critical for enhancing the understanding of individual magnetic component contributions in magnetic assemblages. One of those applicable techniques is isothermal remnant magnetization (IRM) acquisition curves.[16] The IRM acquisition curve method has been used in the environmental and rock magnetism community to advance the understanding of natural processes by studying magnetic mineral assemblages that carry convolved information on environmental mechanisms.[17] It was first suggested by Robertson and France[16] that IRM acquisition curves of a mixed magnetic mineral is a composite curve linearly combined from cumulative log-Gaussian (CLG) functions of individual minerals. Each CLG distribution obtained from IRM acquisition curves represents one individual magnetic phase and allows estimation of its contribution to the bulk magnetization. The raw IRM acquisition curves are usually plotted in 'gradient' obtained by calculating the first derivative of the IRM curve to easily see the distribution of different magnetic phases.[18] In processing a gradient of acquisition plot (GAP),[18] each magnetic component is represented as a log-Gaussian probability density function with a given mean coercivity, dispersion, and relative proportion.

In a typical IRM measurement, the sample should be demagnetized first, then the magnetic field will be applied from zero to a maximum field with a certain step. In order to obtain the remanent data of each magnetic field, the magnetic field will switch back to zero field and the moment at zero field will be recorded as the remanence at this magnetic field. 100 remanence data points are taken in the IRM measurement for Nano-1 at the magnetic field from 0 to 12 kOe.

The experimental IRM acquisition curve for Nano-1 is shown in Fig. S-6(a), and then plotted as a GAP[19] (Fig. S-6(b)) in a coordinate of logarithm of H [log(H)], versus gradient of acquisition, dM/dlog(H).[18] The GAP, which demonstrates the coercivity distribution, reveals that there are three magnetic components with distinct coercivities. The sharp peak in Fig. S-6(b) indicating a Gaussian distribution has a mean coercivity of ~6.5 kOe, which carries the majority of remnant magnetization, and this is consistent with the most intense peak located at ~6.5 kOe in the FORC results (see main paper). Similarly, another peak is fitted at the mean coercivity ~100 Oe, representing the low coercivity phase in the FORC diagram. However, the third fitted peak (green curve) shows a mean coercivity of ~3.2 kOe, which is considerably greater than the peak (~2 kOe) observed in the FORC diagram. The mismatch is presumably due to the coupling between magnetic particles, which affects the successfulness of CLG analysis and, as a result, can suggest misleading interpretations.[16-20] For instance, Heslop et al.[20] found that magnetostatic interaction resulted in left-skewed individual distributions. Since the coercivity distribution of individual components does not always follow the CLG distribution, Egli[21] proposed that this mismatch on peak can be overcome by using skewed generalized Gaussian (SGG) distributions, which enables a more flexible fitting. Therefore, by applying SGG distribution on the second peak, the mean coercivity of it can be modified to a lower magnetic field. This largest gradient obtained at low coercivity phase is presumably due to the imperfect demagnetization process which starts from a very small negative magnetization rather than exactly zero magnetization. This small imperfection is amplified in the low coercivity region. However, the middle peak is now located at a mean coercivity ~2 kOe in Fig. S-6(c) which corresponds to the middle peak as it occurred in the FORC diagram.

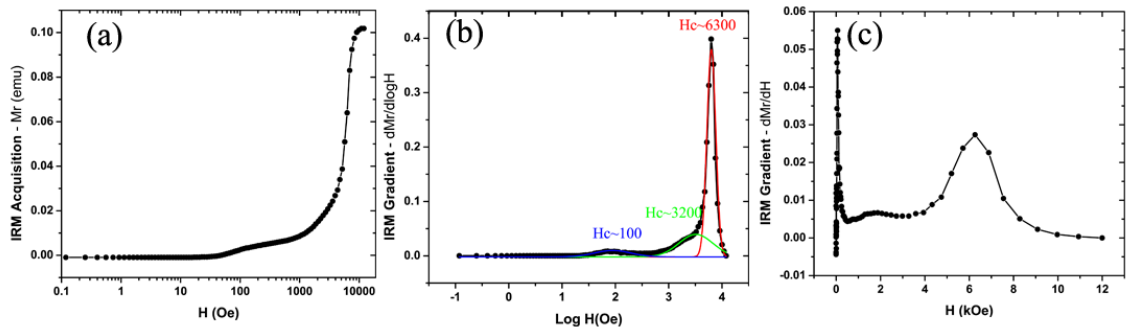

Fig. S-6. Isothermal remnant magnetization (IRM) analysis. (a) IRM acquisition curve of 'barium hexaferrite -lot 1' and the gradient curves in (b) log scale and (c) linear scale. Three cumulative log Gaussian distribution functions were used to fit the gradient curve in log scale.

C. FORC data collections

Four commercial samples purchased as M-type barium hexaferrite ($BaFe_{12}O_{19}$) had FORCs measured multiple times at Washington State University (WSU) or Lakeshore. All parameters of each FORC measurement are listed in Table S-IV.

**Table S-IV**. FORC parameters. $H_{u1}$ and $H_{u2}$ are the limits for the y-axis (interaction or bias) on the FORC diagram. $H_{c1}$ and $H_{c2}$ are the limits for the x-axis (coercivity axis) on the FORC diagram. $H_{Cal}$ and HSat are the fields used for calibration or saturation, respectively, applied at the end of each FORC. $H_{Ncr}$ is field step between reversal fields and $N_{Forc}$ is the number of FORCs.

| Sample | $H_{u1}$ (kOe) | $H_{u2}$ (kOe) | $H_{c1}$ (kOe) | $H_{c2}$ (kOe) | $H_{Cal}$ (kOe) | $H_{Sat}$ (kOe) | $H_{Ncr}$ (Oe) | $N_{Forc}$ | Date | Figure # |
|---|---|---|---|---|---|---|---|---|---|---|
| Micro | -1.5 | 1.5 | 0 | 10 | 11.8 | 20 | 59.65 | 150 | 1/5/2017 | i |
| Nano-1 | -1.5 | 1.5 | 0 | 10 | 11.8 | 20 | 208.84 | 150 | 1/3/2017 | ii |
| Nano-1 | -1.5 | 1.5 | 0 | 10 | 11.5 | 32 | 131.34 | 100 | 6/22/2016 | iii |
| Nano-1 | -1.5 | 1.5 | 0 | 28 | 29.5 | 33 | 208.84 | 150 | 6/22/2016 | iv |
| Nano-2 | -1 | 1 | 0 | 1.5 | 2.5 | 32 | 35.15 | 100 | 6/23/2016 | v |
| Nano-2 | -1.5 | 1.5 | 0 | 28 | 29.5 | 33 | 208.84 | 150 | 1/23/2017 | vi |
| Nano-3 | -1.5 | 1.5 | 0 | 1 | 2.5 | 32 | 40.08 | 100 | 6/22/2016 | vii |
| Nano-3 | -1.5 | 1.5 | 0 | 2 | 3.72 | 10 | 52.6 | 100 | 1/5/2017 | viii |
| Nano-3 | -0.5 | 0.5 | 0 | 10 | 10.5 | 28 | 73.31 | 150 | 7/24/2017 | ix |
| Nano-3 | -1.5 | 1.5 | 0 | 28 | 29.5 | 33 | 208.84 | 150 | 12/11/2017 | x |

On the following pages, iterative FORC and FORC diagram examples are given using the different parameters shown in the Table above. The final FORC pattern in the main paper are shown for completeness. Note that these presented diagrams do not have the lower edge artifact[13] removed during data processing.

i. Sample: Ba-hex-micro

Date: 1/5/2017 (Lakeshore PMC 3900 VSM) – NOTE: this is the same data presented in the main paper

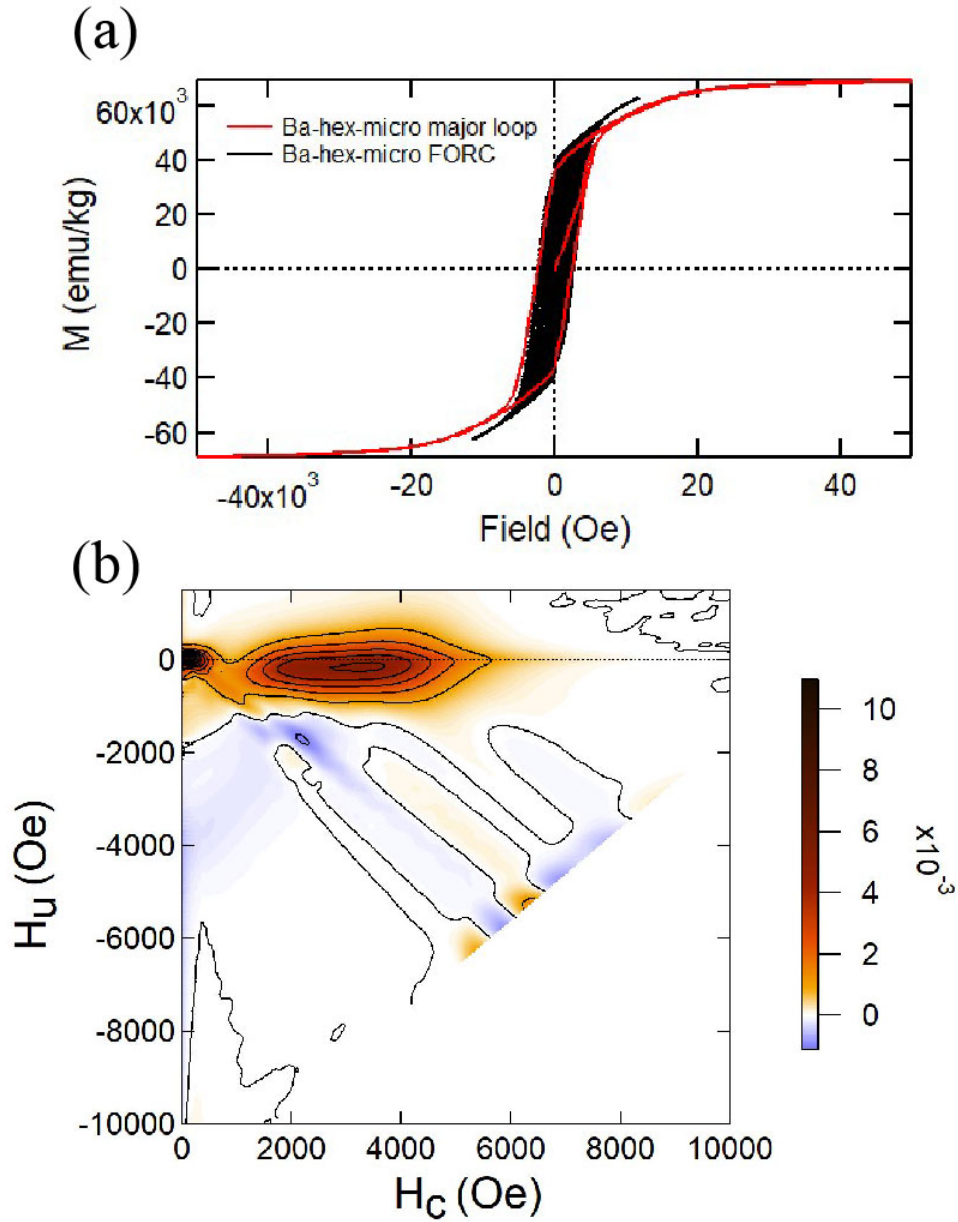

- Measurement parameters:
  Hb1             -1.500000E+03
  Hb2             +1.500000E+03
  Hc1              0.000000E+00
  Hc2             +10.00000E+03
  HCal            +11.84507E+03

ii. Sample: Ba-hex-nano-1-WSU

Date: 1/3/2017 (Lakeshore PMC 3900 VSM, WSU)

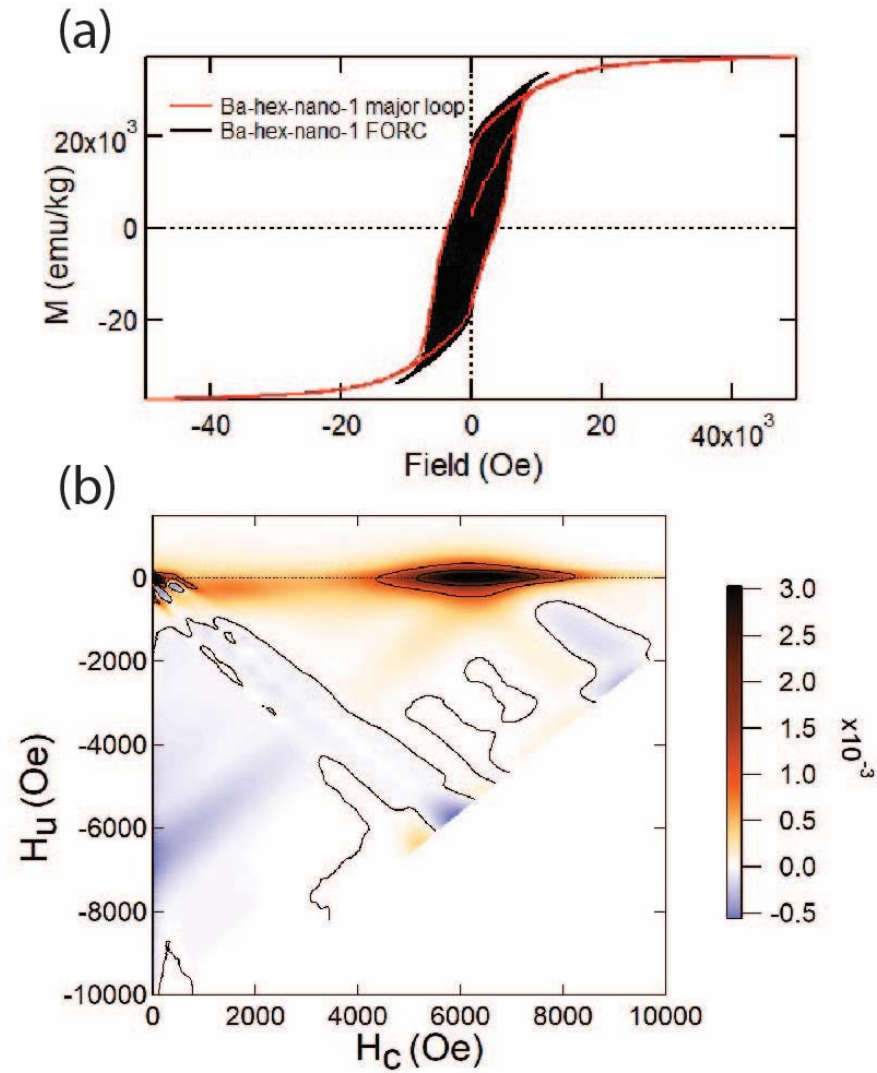

- Measurement parameters:

| | |
|---|---|
| Hb1 | -1.500000E+03 |
| Hb2 | +1.500000E+03 |
| Hc1 | 0.000000E+00 |
| Hc2 | +10.00000E+03 |
| HCal | +11.84507E+03 |
| HNcr | +89.65527E+00 |
| HSat | +20.00000E+03 |
| NForc | 150 |

iii. Sample: Ba-hex-nano-1-10 kOe

Date: 6/22/2016 (Lakeshore PMC 3900 VSM)

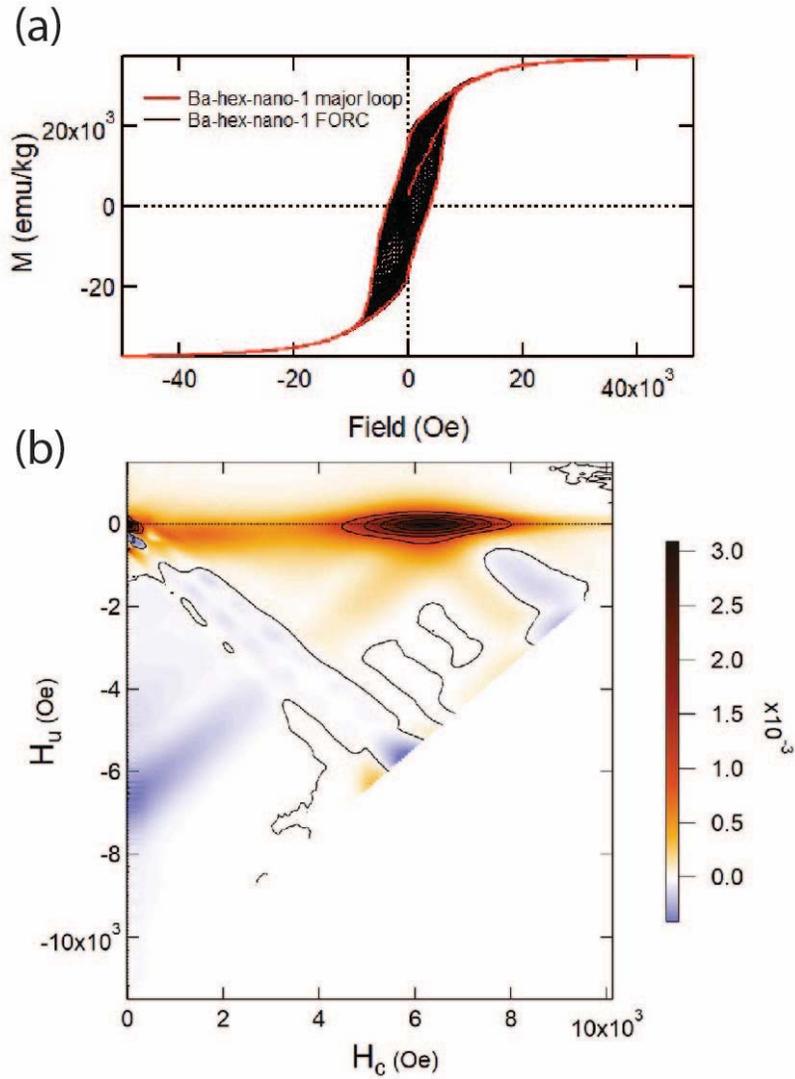

- Measurement parameters:
  Hb1           -1.50000E3
  Hb2           1.50000E3
  Hc2           10.0000E3
  HCal          11.5000E3
  HSat          32.0000E3
  NForc         100

iv. Sample: Ba-hex-nano-1 -28 KOe

Date: 12/12/2017 (Lakeshore 7400 VSM) – NOTE: this is the same data presented in the main paper

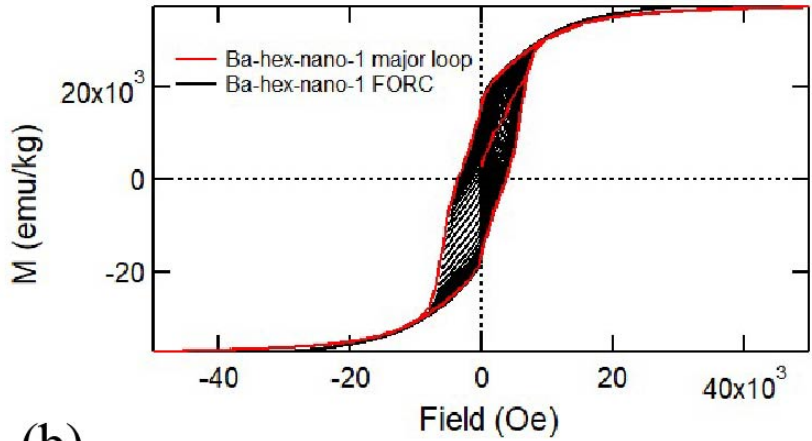

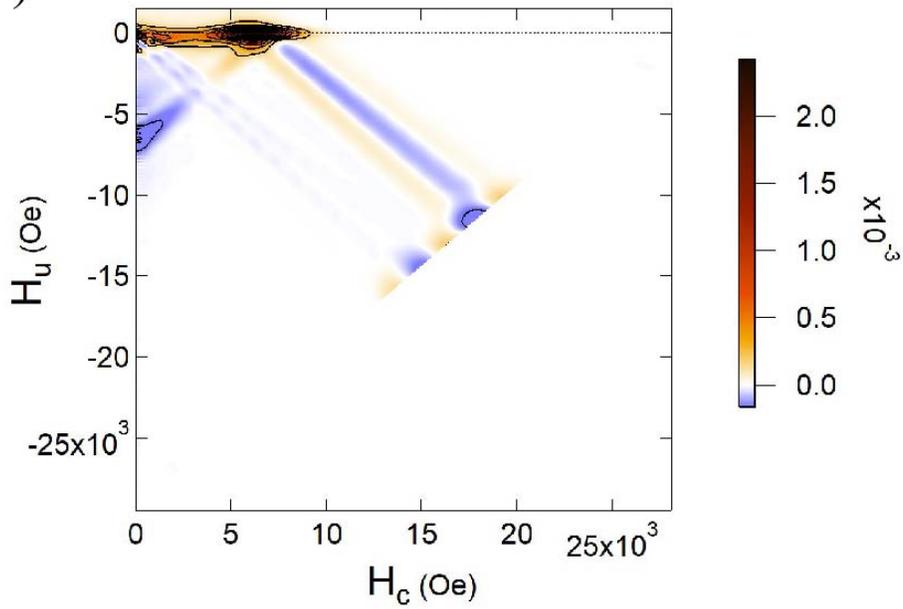

- Measurement parameters:
  | | |
  |---|---|
  | Hb1 | -1.50000E3 |
  | Hb2 | 1.50000E3 |
  | Hc2 | 28.0000E3 |
  | HCal | 29.5000E3 |
  | HSat | 33.0000E3 |
  | NForc | 150 |

v. Sample: Ba-hex-nano-2-1.5 kOe

Date: 6/23/2016 (Lakeshore PMC 3900 VSM)

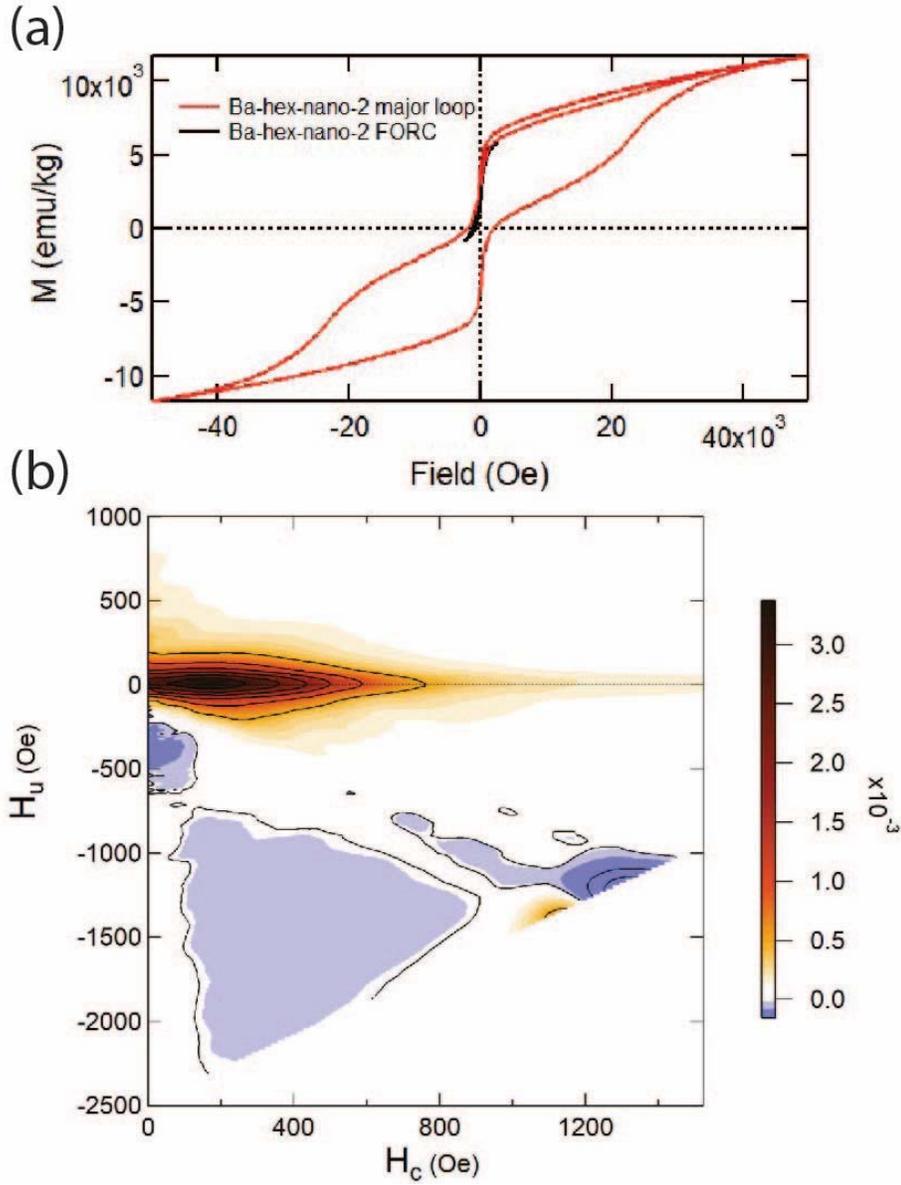

- Measurement parameters:
  | | |
  |---|---|
  | Hb1 | -1.00000E3 |
  | Hb2 | 1.00000E3 |
  | Hc2 | 1.50000E3 |
  | HCal | 2.50000E3 |
  | HSat | 32.0000E3 |
  | NForc | 100 |

vi. Sample: Ba-hex-nano-2-28 kOe

Date: 1/23/2017 (Lakeshore 7400 VSM) – NOTE: this is the same data presented in the main paper

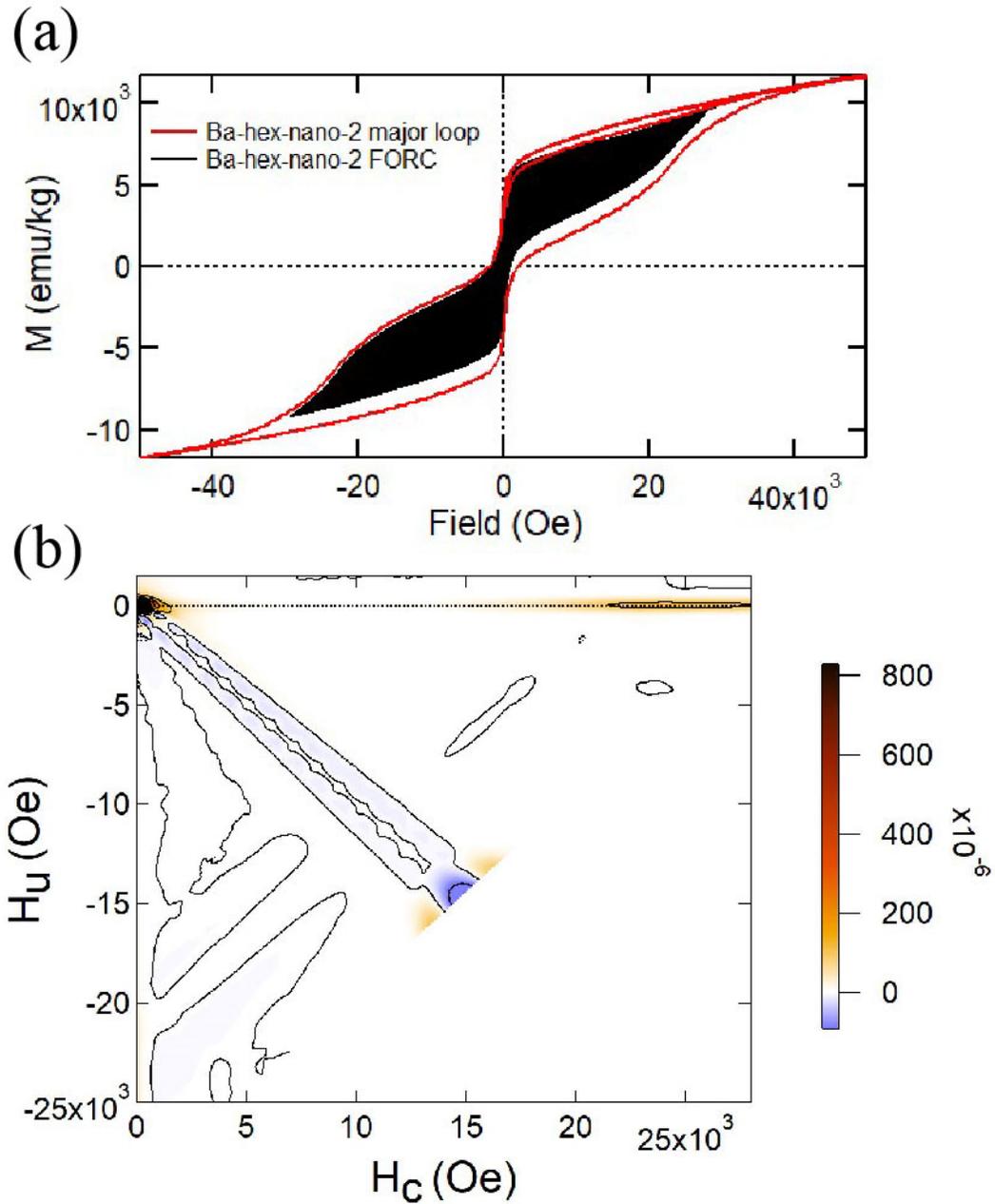

- Measurement parameters:
  - Hb1     -1.50000E3
  - Hb2     1.50000E3
  - Hc2     28.0000E3
  - HCal    29.5000E3
  - HSat    33.0000E3
  - NForc   150

vii. Sample: Ba-hex-nano-3-Lakeshore-1 kOe

Date: 6/22/2016 (Lakeshore PMC 3900 VSM)

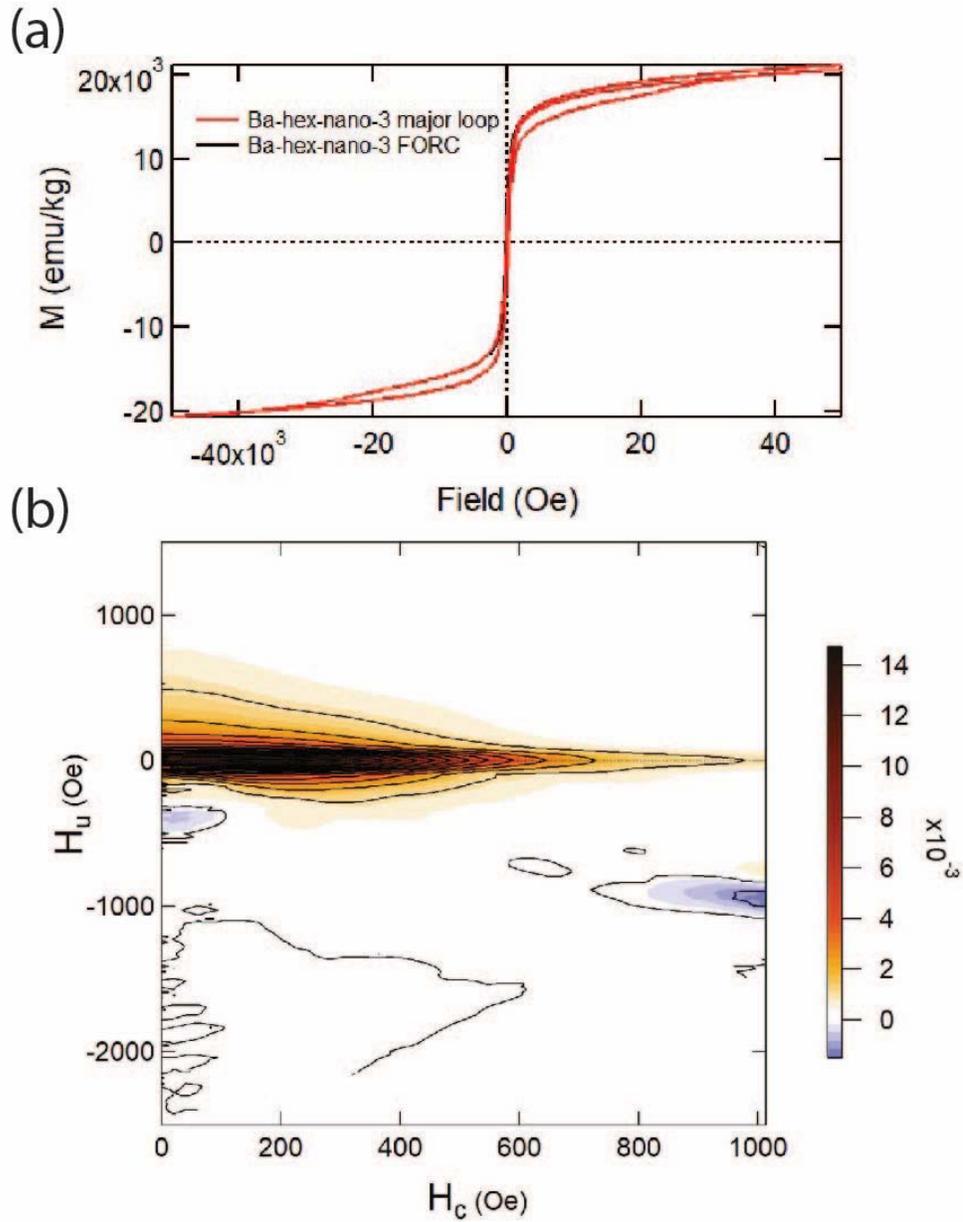

- Measurement parameters:
    Hb1         -1.50000E3
    Hb2          1.50000E3
    Hc2          1.00000E3
    HCal         2.50000E3
    HSat         32.0000E3
    NForc        100

viii. Sample: Ba-hex-nano-3-2 kOe

Date: 1/5/2017 (Lakeshore PMC 3900 VSM)

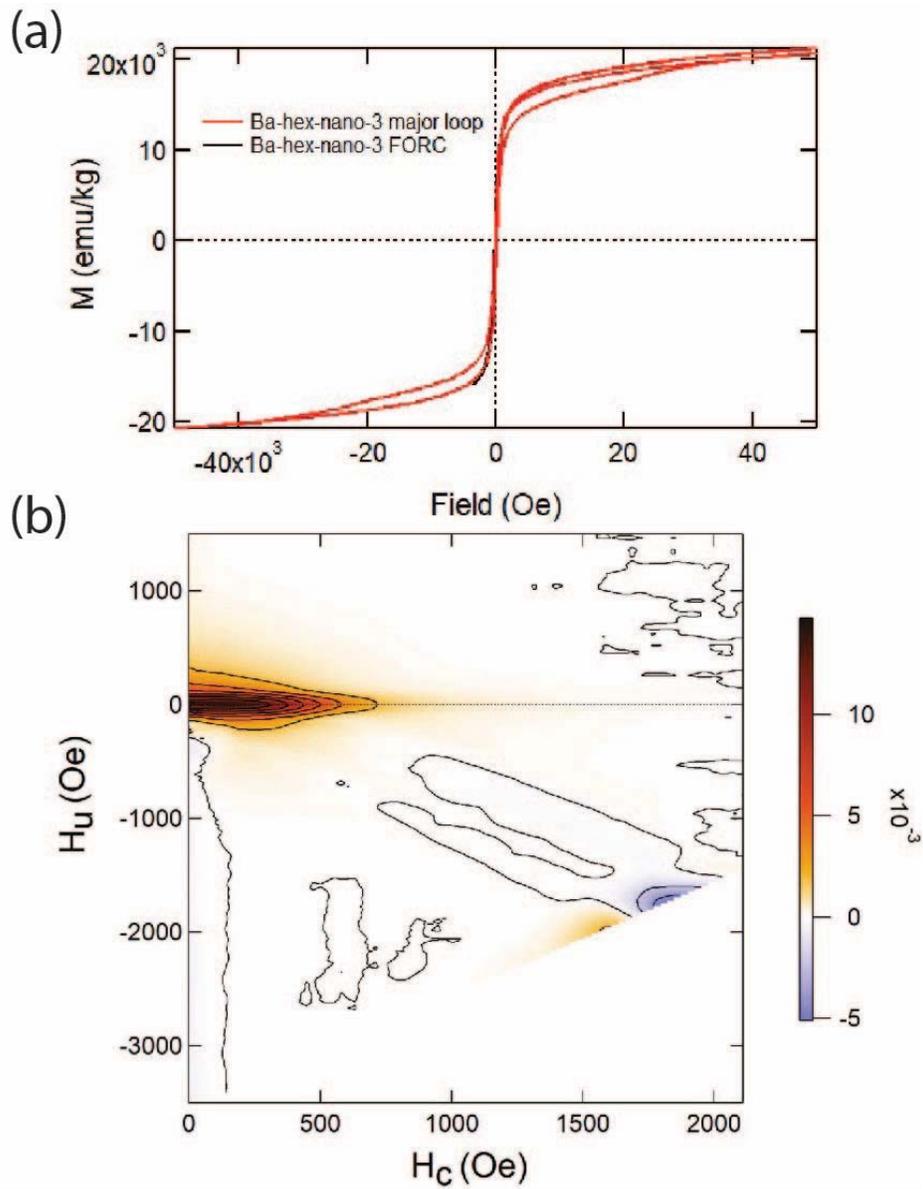

- Measurement parameters:

  | | |
  |---|---|
  | Hb1 | -1.500000E+03 |
  | Hb2 | +1.500000E+03 |
  | Hc1 | 0.000000E+00 |
  | Hc2 | +2.000000E+03 |
  | HCal | +3.726484E+03 |
  | HNcr | +52.63163E+00 |
  | HSat | +10.00000E+03 |
  | NForc | 100 |

ix. Sample: Ba-hex-nano-3-10 kOe

Date: 7/24/2017 (Lakeshore PMC 3900 VSM)

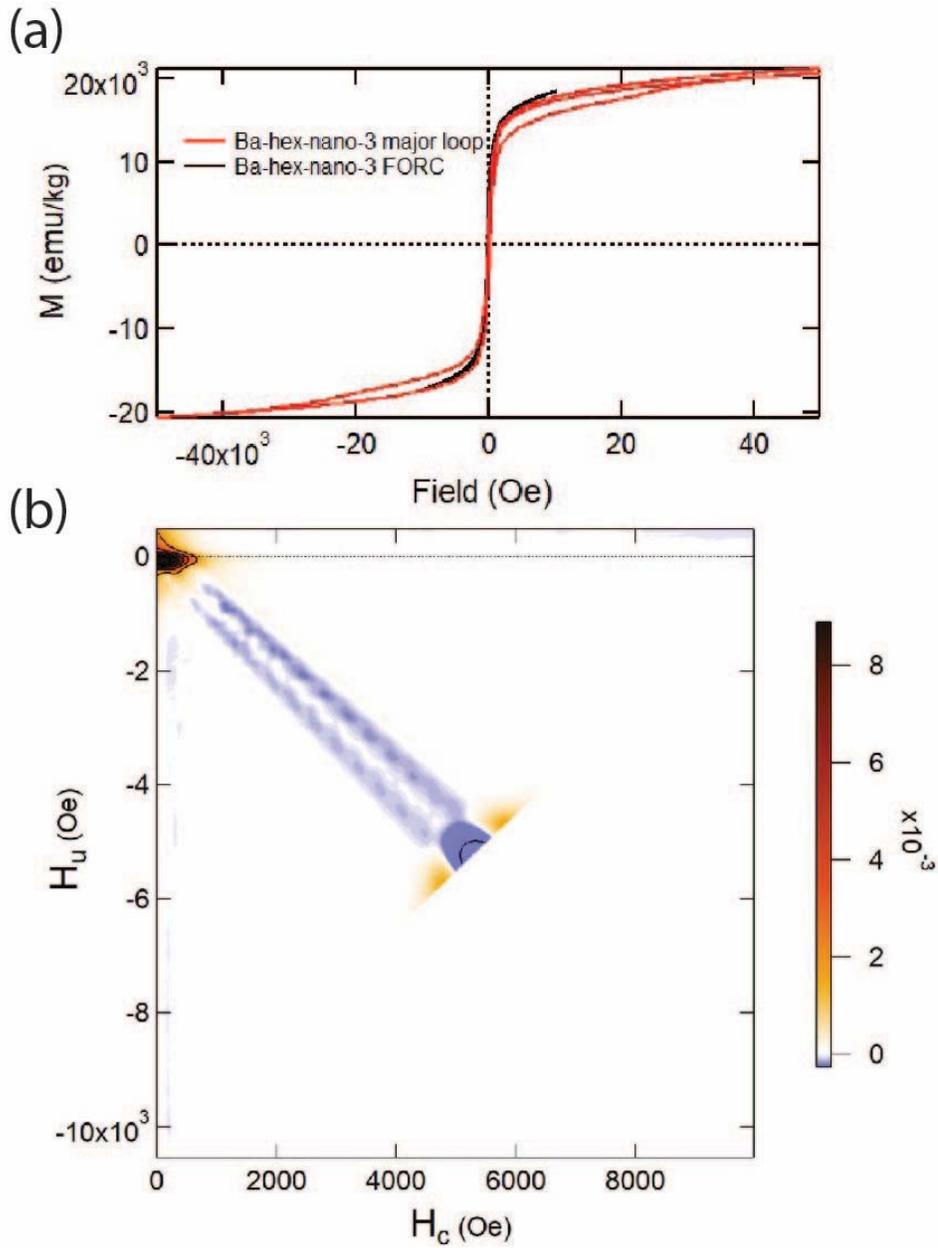

- Measurement parameters:
  | | |
  |---|---|
  | Hb1 | -500 |
  | Hb2 | 500 |
  | Hc2 | 10000 |
  | HCal | 10500 |
  | HSat | 28000 |
  | NForc | 150 |

x. Sample: Ba-hex-nano-3-28 kOe

Date: 12/11/2017 (Lakeshore 7400 VSM) – NOTE: this is the same data presented in the main paper

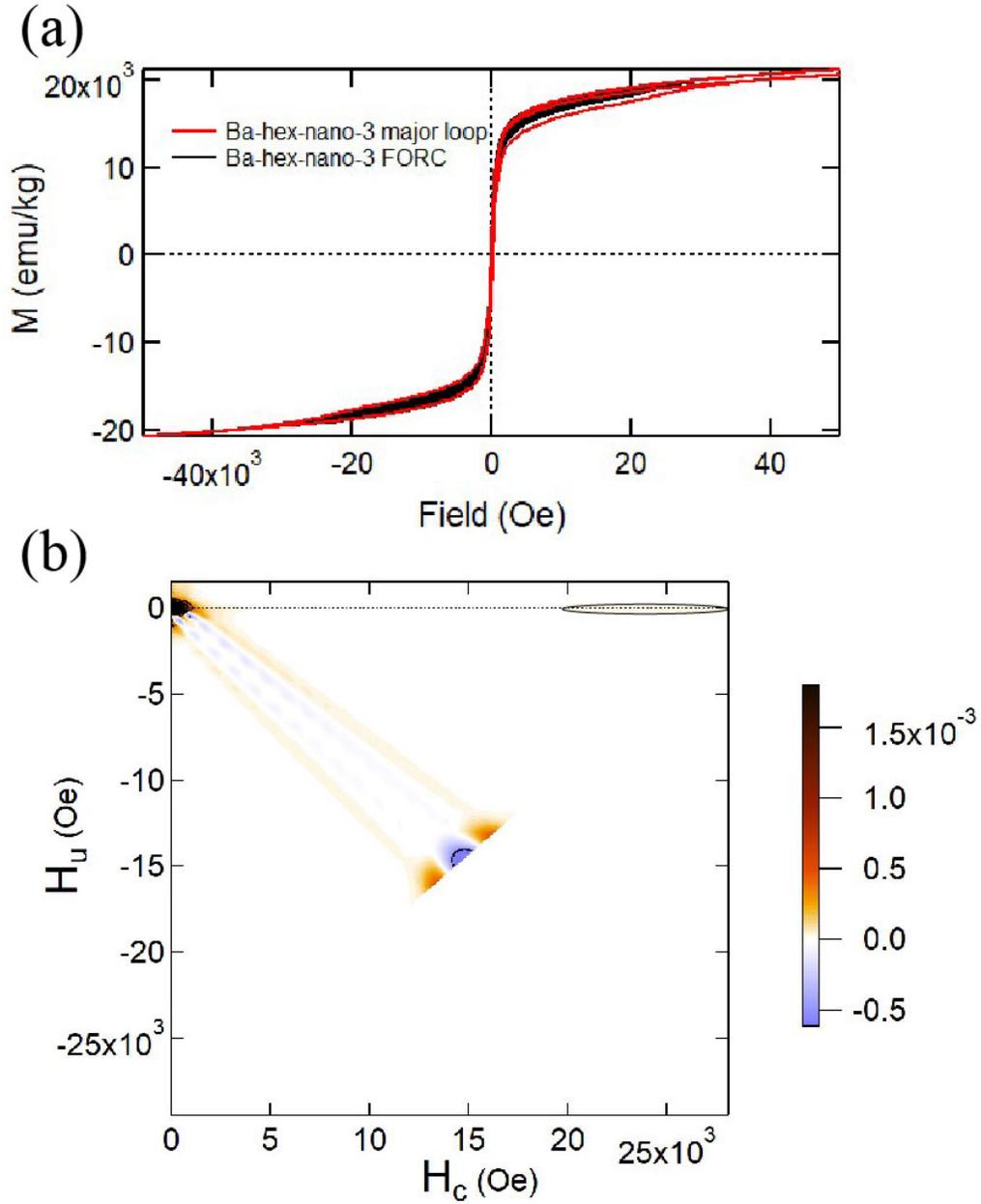

- Measurement parameters:
  Hb1         -1.50000E3
  Hb2         1.50000E3
  Hc2         28.0000E3
  HCal        29.5000E3
  HSat        33.0000E3
  NForc       150